\documentstyle[tighten,preprint,amsfonts,%
eqsecnum,aps,prd,epsf]{revtex}

\begin{document}
\draft

\hyphenation{
mani-fold
mani-folds
geo-metry
geo-met-ric
}

%% MACROS

%% uncomment the first two lines below if
%% amsfonts is not available
%%
%\def\Bbb{\bf}
%\def\frak{\bf}

\def\BbbR{{\Bbb R}}
\def\BbbZ{{\Bbb Z}}
\def\BbbC{{\Bbb C}}

\def\RPthree{{{\Bbb RP}^3}}
\def\RPtwo{{{\Bbb RP}^2}}

\def\casehalf{{\case{1}{2}}}

\def\minkvac{{|0\rangle}}
\def\noughtvac{{|0_0\rangle}}
\def\minusvac{{|0_-\rangle}}

\def\minkvacbra{{\langle0|}}
\def\noughtvacbra{{\langle 0_0|}}
\def\minusvacbra{{\langle 0_-|}}

\def\rindvacminus{{|0_{R_-}\rangle}}

\def\boulvac{{|0_{\rm B}\rangle}}

\def\hhvackrus{{|0_{\rm K}\rangle}}
\def\hhvacgeon{{|0_{\rm G}\rangle}}

\def\hhvacgeonbra{{\langle 0_{\rm G}|}}
\def\hhvackrusbra{{\langle 0_{\rm K}|}}

\def\ttilde{{\tilde t}}
\def\etatilde{{\tilde \eta}}

\def\fieldmass{{\mu}}

\def\Othree{{\rm O}(3)}
\def\SOthree{{\rm SO}(3)}

\def\Uone{{\rm U}(1)}

\def\arcsinh{\mathop{\rm arcsinh}\nolimits}

%% PAPER BEGINS

\preprint{\vbox{\baselineskip=12pt
\rightline{PPG98-79}
\rightline{SU-GP-98/2-1}
\rightline{gr-qc/9802068}}}
\title{Inextendible Schwarzschild black hole 
with a single exterior: 
\\
How thermal is the Hawking radiation?}
\author{Jorma Louko\footnote{%
Electronic address:
louko@aei-potsdam.mpg.de
}}
\address{
Department of Physics,
University of Maryland,
College Park,
Maryland 20742--4111,
USA\\
and\\
Max-Planck-Institut f\"ur Gravitations\-physik,
Schlaatzweg~1,
D--14473 Potsdam,
Germany\footnote{Present address.}
}
\author{Donald Marolf\footnote{Electronic address:
marolf@suhep.phy.syr.edu}}
\address{
Department of Physics,
Syracuse University,
Syracuse, New York 13244--1130, USA}
\date{Published in 
{\it Phys.\ Rev.\ \rm D \bf 58}, 024007 (1998)}
\maketitle
\begin{abstract}%
Several approaches to Hawking radiation on Schwarzschild spacetime
rely in some way or another on the fact that the Kruskal manifold has
two causally disconnected exterior regions. To assess the physical
input implied by the presence of the second exterior region, we
investigate the Hawking(-Unruh) effect for a real scalar field on the
$\RPthree$ geon: an inextendible, globally hyperbolic, 
space and time orientable eternal black hole spacetime that is 
locally isometric to Kruskal but contains only one exterior region. 
The Hartle-Hawking-like vacuum~$\hhvacgeon$,
which can be characterized alternatively by the positive frequency
properties along the horizons or by the complex analytic properties of
the Feynman propagator, turns out to contain exterior region Boulware
modes in correlated pairs, and any operator in the exterior that only
couples to one member of each correlated Boulware pair has thermal
expectation values in the usual Hawking temperature. Generic operators
in the exterior do not have this special form; however, we use a
Bogoliubov transformation, a particle detector analysis, and a
particle emission-absorption analysis that invokes the analytic
properties of the Feynman propagator, to argue that $\hhvacgeon$
appears as a thermal bath with the standard Hawking temperature to any
exterior observer at asymptotically early and late Schwarzschild
times. A~(naive) saddle-point estimate for the path-integral-approach
partition function yields for the geon only half of the
Bekenstein-Hawking entropy of a Schwarzschild black hole with the same
ADM mass: possible implications of this result for the validity of
path-integral methods or for the statistical interpretation of
black-hole entropy are discussed. Analogous results hold for a Rindler
observer in a flat spacetime whose global properties mimic those of
the geon.
\end{abstract}

\narrowtext

\section{Introduction}
\label{sec:intro}

Black hole entropy was first put on a firm footing by combining
Hawking's result of black hole radiation \cite{hawkingCMP} with the
dynamical laws of classical black hole geometries \cite{BarCarHaw} in
the manner anticipated by Bekenstein \cite{bekenstein1,bekenstein2}.
Hawking's first calculation of black hole temperature
\cite{hawkingCMP} invoked quantum field theory in a time-nonsymmetric
spacetime that modeled a collapsing star, and the resulting
time-nonsymmetric quantum state contained a net flux of radiation
from the black hole \cite{unruh-magnum}. However, it was soon
realized that the same temperature, and hence the same entropy, is
also associated with
a time-symmetric state that describes a thermal
equilibrium \cite{hh-vacuum,israel-vacuum}. For a review, see for
example Ref.\ \cite{wald-qft}.

A second avenue to black hole entropy has arisen via path integral
methods. Here, a judiciously chosen set of thermodynamic variables
is translated into geometrical boundary conditions for a
gravitational path integral,
and the path integral is then interpreted as a partition function in
the appropriate thermodynamic ensemble. The initial impetus for the
path-integral approach came in the observation \cite{GH1,hawkingCC}
that for the Kerr-Newman family of black holes in asymptotically flat
space, a saddle-point estimate of the path integral yields a
partition function that reproduces the Bekenstein-Hawking black hole
entropy. The subject has since evolved considerably; see for example
Refs.\ \cite{pagerev,BY-quasilocal,BY-microcan,%
HawHoroRoss,teitel-extremal}, and the references therein.

Although it is empirically true that these two methods for arriving
at black hole entropy have given mutually compatible results in
most\footnote{For a discussion of discrepancies for extremal holes,
see Refs.\ \cite{HawHoroRoss,teitel-extremal}.}
situations considered, it does not seem to be well understood why
this should be the case. The first method is quite indirect, and it
gives few hints as to the quantum gravitational degrees of freedom
that presumably underlie the black hole entropy. In contrast, the
path integrals of the second method arise from quantum gravity
proper, but the argument is quite formal,
and one is left with the challenge of
justifying that the boundary conditions imposed on these integrals
indeed correspond to thermodynamics as conventionally understood. One
expects that the connection between the path integrals and the
thermodynamics could be made precise through some appropriate
operator formulation, as is the case in Minkowski space finite
temperature field theory \cite{kapusta}. Achieving such an operator
formulation in quantum gravity does however not appear imminent, the
recent progress in string theory \cite{polchinski-coll,horo-chandra}
notwithstanding.

The purpose of this paper is to examine the Hawking effect and
gravitational entropy on the eternal black hole spacetime known as the
$\RPthree$ geon \cite{topocen}. This inextendible vacuum Einstein
spacetime is locally isometric to the Kruskal manifold, and it in
particular contains one exterior Schwarzschild region. The spacetime
is also both space and time orientable and globally hyperbolic, and
hence free of any apparent pathologies. A novel feature is, however,
that the black and white hole interior regions are not globally
isometric to those of the Kruskal manifold. Also, there is no second
exterior Schwarzschild region, and the timelike Killing vector of the
single exterior Schwarzschild region cannot be extended into a
globally-defined Killing vector on the whole spacetime. Among the
continuum of constant Schwarzschild time hypersurfaces in the exterior
region, there is only one that can be extended into a smooth Cauchy
hypersurface for the whole spacetime, but probing only the exterior
region provides no clue as to which of the constant Schwarzschild time
hypersurfaces this one actually is.\footnote{Another inextendible
  spacetime that is locally isometric to Kruskal
  but contains only one exterior Schwarzschild region is the
  elliptic interpretation of the Schwarzschild hole,
  investigated in Ref.\ \cite{gibbons-elliptic} in the context of
  't~Hooft's analysis of Hawking radiation \cite{tHooft}. On this
  spacetime, all the local continuous isometries can be extended into
  global ones. The spacetime is, however, not time-orientable, which
  gives rise to subtleties when one wishes to build a quantum field
  theory with a Fock space \cite{gibbons-elliptic}.}

These features of the $\RPthree$ geon lead one to ask to what
extent quantum physics on this spacetime, especially in the
exterior region, knows that the spacetime differs from Kruskal
behind the horizons. In particular, is there a Hawking effect, and
if yes, can an observer in the exterior region distinguish this
Hawking effect from that on the Kruskal manifold? Also, can one
attribute to the $\RPthree$ geon a gravitational entropy by either
of the two methods mentioned above, and if yes, does this entropy
agree with that for the Kruskal spacetime?

Answers to these questions have to start with the specification of
the quantum state of the field(s) on the $\RPthree$ geon. To this
end, we recall that the geon can be constructed as the quotient
space of the Kruskal manifold under an involutive isometry
\cite{topocen}. Any vacuum on Kruskal that is invariant under this
involution therefore induces a vacuum on the geon. This is in
particular the case for the Hartle-Hawking vacuum $\hhvackrus$
\cite{hh-vacuum,israel-vacuum}, which describes a Kruskal hole in
equilibrium with a thermal bath at the Hawking temperature $T =
{(8\pi M)}^{-1}$. We shall fix our attention to the
Hartle-Hawking-like vacuum $\hhvacgeon$ that
$\hhvackrus$ induces on the geon. $\hhvacgeon$~can alternatively
be defined (see subsection~\ref{subsec:boul-hh-bogo})
by postulating for its Feynman propagator a suitable relation with
Green's functions on the Riemannian section of the complexified
manifold, in analogy with the path-integral derivation of
$\hhvackrus$ in Ref.\ \cite{hh-vacuum}. A~final definition which
leads to the same vacuum state is to construct $\hhvacgeon$ as the
state defined by modes that are positive frequency along the
horizon generators.

We first construct the Bogoliubov transformation between
$\hhvacgeon$ and the Boulware vacuum~$\boulvac$, which is the
vacuum with respect to the timelike Killing vector of the exterior
Schwarzschild region \cite{boulware-vac,fulling}. For a massless
scalar field, we find that $\hhvacgeon$ contains Boulware modes in
correlated pairs, and for operators that only couple to one member
of each correlated pair, the expectation values in $\hhvacgeon$
are given by a thermal density matrix at the usual Hawking
temperature. As both members of each correlated pair
reside in the single exterior Schwarzschild region, not every
operator with support in the exterior region has this particular
form; nevertheless, we find that, far from the black hole, this 
{\em is\/}
the form assumed by every
operator whose support is at asymptotically late (or early) values
of the exterior Schwarzschild time. For a massive scalar field,
similar statements hold for the field modes that reach the
infinity. As a side result, we obtain an explicit demonstration
that the restriction of $\hhvacgeon$ to the exterior region is not
invariant under translations in the Schwarzschild time.

The contrast between these results and those in the vacuum
$\hhvackrus$ on the Kruskal manifold \cite{israel-vacuum} is clear.
$\hhvackrus$~is also a superposition of correlated pairs of
Boulware modes, but the members of each correlated pair in
$\hhvackrus$ reside in the opposite exterior Schwarzschild regions
of the Kruskal manifold. In~$\hhvackrus$, the expectation values
are thermal for any operators with support in just one of the two
exterior Schwarzschild regions.

We then consider the response of a monopole particle detector
\cite{unruh-magnum,dewitt-CC,birrell-davies,takagi-magnum} in the
vacuum~$\hhvacgeon$. The detector is taken to be in the exterior
Schwarzschild region, and static with respect to the Schwarzschild
time translation Killing vector of this region. The response turns out
to differ from that of a similar detector in the vacuum $\hhvackrus$
on Kruskal; in particular, while the response on Kruskal is static,
the response on the geon is not. However, we argue that the responses
on the geon and on Kruskal should become identical in the limit of
early or late geon Schwarzschild times (as might be inferred from the
Bogoliubov transformation described above) and also in the limit of a
detector at large curvature radius for any fixed geon Schwarzschild
time. To make the argument rigorous, it would be sufficient to verify
certain technical assumptions about the falloff of the Wightman
function $G^+$ in~$\hhvackrus$.

We proceed to examine the complex analytic properties of the
Feynman propagator $G_{\rm G}^F$ in~$\hhvacgeon$. The quotient
construction of the geon from the Lorentzian Kruskal manifold can
be analytically continued, via the formalism of (anti)holomorphic
involutions on the complexified manifolds
\cite{gibb-holo,chamb-gibb}, into a quotient construction of the
Riemannian section of the geon from the Riemannian Kruskal
manifold. It follows that $G_{\rm G}^F$ is regular on the
Riemannian section of the geon everywhere except at the coincidence
limit. $G_{\rm G}^F$~turns out to be, in a certain weak local
sense, periodic in the Riemannian Schwarzschild time with period
$8\pi M$ in each argument. However, this local periodicity is not
associated with a continuous invariance under simultaneous
translations of both arguments in the Riemannian Schwarzschild
time. Put differently, the Riemannian section of the geon does not
admit a globally-defined Killing vector that would locally coincide
with a generator of 
translations in the Riemannian Schwarzschild time. It is
therefore not obvious what to conclude about the thermality of
$\hhvacgeon$ by just inspecting the symmetries of $G_{\rm G}^F$ on
the Riemannian section.  Nevertheless,
we can use the analytic properties
of $G_{\rm G}^F$ to relate the probabilities of the geon to emit
and absorb a Boulware particle with a given frequency, in analogy
with the calculation done for the Kruskal spacetime in Ref.\
\cite{hh-vacuum}. We find that the probability for the geon to emit
a particle with frequency $\omega$ at late exterior Schwarzschild
times is $e^{-8\pi M\omega}$ times the probability for the geon to
absorb a particle in the same mode. This ratio of the probabilities
is characteristic of a thermal spectrum at the Hawking temperature
$T = {(8\pi M)}^{-1}$, and it agrees with that obtained for
$\hhvackrus$ in Ref.\ \cite{hh-vacuum}. A~difference between
Kruskal and the geon is, however, that the Killing time translation
isometry of the Kruskal manifold guarantees the thermal result for
$\hhvackrus$ to hold for particles at arbitrary values of the
exterior Schwarzschild time, while we have not been able to relax
the assumption of late exterior Schwarzschild times
for~$\hhvacgeon$.

These results for the thermal properties of $\hhvacgeon$ imply that an
observer in the exterior region of the geon, at late Schwarzschild
times, can promote the classical first law of black hole mechanics
into a first law of black hole thermodynamics exactly as for the
Kruskal black hole. Such an observer thus finds for the thermodynamic
entropy of the geon the usual Kruskal value~$4 \pi M^2$, which is one
quarter of the area of the geon black hole horizon at late times.  If
one views the geon as a dynamical black-hole spacetime, with the
asymptotic far-future horizon area~$16 \pi M^2$, this is the result
one might have expected on physical grounds.

On the other hand, the area-entropy relation for the geon is made
subtle by the fact that the horizon area is in fact not constant
along the horizon. Away from the intersection of the past and
future horizons, the horizon duly has topology $S^2$ and area $16
\pi M^2$, just as in Kruskal. The critical surface at the
intersection of the past and future horizons, however,
has topology $\RPtwo$ and area~$8 \pi M^2$. As it is precisely this
critical surface that belongs to both the Lorentzian and Riemannian
sections of the complexified manifold, and constitutes the horizon
of the Riemannian section, one may expect that methods utilizing the
analytic structure of the geon and the Riemannian section of the
complexified manifold would produce for the entropy the value~$2
\pi M^2$, which is one quarter of the critical surface area, and
only half of the Kruskal entropy. We shall find that this is indeed
the semiclassical geon entropy that emerges from the path-integral
formalism, when the boundary conditions for the path integral are
chosen so that the saddle point is the Riemannian section of the
geon.

Several viewpoints on this discrepancy between the thermodynamic late
time entropy and the path-integral entropy are possible. At one
extreme, there are reasonable grounds to suspect outright the
applicability of the path-integral methods to the geon. At another
extreme, the path-integral entropy might be correct but physically
distinct from the subjective thermodynamic entropy seen by a late
time exterior observer. For example, a physical interpretation for
the path-integral entropy might be sought in the quantum statistics
in the whole exterior region, rather than just in the thermodynamics
at late times in the exterior region.

All these results for the geon turn out to have close
counterparts in the thermodynamics of an accelerated observer in a
flat spacetime $M_-$ whose global properties mimic those of the
geon. $M_-$~has a global timelike Killing vector that
defines a Minkowski-like vacuum~$\minusvac$, but it has only one
Rindler wedge, and the Rindler time translations in this wedge cannot
be extended into globally-defined isometries of~$M_-$. $\minusvac$~is
thus analogous to the Hartle-Hawking-like vacuum $\hhvacgeon$ on the
geon, and the Rindler vacuum in the Rindler wedge of $M_-$ is
analogous to the Boulware vacuum~$\boulvac$. We find, from a
Bogoliubov transformation, a particle detector calculation, and the
analytic properties of the Feynman propagator, that the accelerated
observer sees $\minusvac$ as a thermal bath at the Rindler
temperature under a restricted class of observations, and in
particular in the limit of early and late Rindler times, but not
under all observations.
Note, however, that $M_-$ does not exhibit a nontrivial analogue of
the large curvature radius limit of the geon. The reason for this is
that $\minusvac$ and the Rindler vacuum coincide far from the
acceleration horizon, just as the
Minkowski-vacuum and the usual
Rindler-vacuum coincide far from the acceleration horizon in the
topologically trivial case. 

For a massless field, 
we also compute the renormalized expectation value of
the stress-energy tensor in~$\minusvac$. This expectation value is
not invariant under Rindler time translations in the Rindler wedge,
but the noninvariant piece turns out to vanish
in the limit of early and late Rindler times, as well as in the limit
of large distances from the acceleration horizon.
Results concerning the entropy of flat spaces \cite{laf-flat-entropy}
are again similar to those mentioned above for the geon entropy.

The rest of the paper is as follows. Sections \ref{sec:flats} and
\ref{sec:observer-on-flats} are devoted to the accelerated observer
on~$M_-$: section \ref{sec:flats} constructs the Minkowski-like vacuum
and finds the renormalized expectation value of the stress-energy
tensor, while section \ref{sec:observer-on-flats} analyzes the
Bogoliubov transformation in the Rindler wedge, a particle detector,
and the analytic properties of the Feynman propagator.  Section
\ref{sec:kruskal-rpthree-defs} is a mathematical interlude in which we
describe the complexified $\RPthree$ geon manifold as a quotient space
of the complexified Kruskal manifold with respect to an holomorphic
involution: this formalizes the sense in which the Riemannian section
of the geon can be regarded as a quotient space of the Riemannian
Kruskal manifold.  Section \ref{sec:field-on-rpthree} analyzes the
vacuum $\hhvacgeon$ in terms of a Bogoliubov transformation, a
particle detector, and the analytic properties of the Feynman
propagator.  Section \ref{sec:rpthree-entropy} addresses the entropy
of the geon from both the thermodynamic and path integral points of
view, and discusses the results in light of the previous sections.
Finally, section \ref{sec:discussion} summarizes the results and
discusses remaining issues.

We work in Planck units, $\hbar = c = G = 1$. A metric with signature
$({-}{+}{+}{+})$ is called Lorentzian, and a metric with signature
$({+}{+}{+}{+})$ Riemannian. All scalar fields are global sections of
a real line bundle over the spacetime ({\em i.e.,} we do not consider
twisted fields). Complex conjugation is denoted by an overline.

A~note on the terminology is in order. The name ``Hawking effect" is
sometimes reserved for particle production in a collapsing star
spacetime, while the existence of a thermal equilibrium state in a
spacetime with a bifurcate Killing horizon is referred to as the Unruh
effect; see for example Ref.\ \cite{wald-qft}. In this terminology,
the partial thermal properties of $\hhvacgeon$ and $\minusvac$ might
most naturally be called a generalized Unruh effect, as these states
are induced by genuine Unruh effect states on the double cover
spacetimes.  However, neither the geon nor $M_-$ in fact has a
bifurcate Killing horizon, and our case study seems not yet to
establish the larger geometrical context of the thermal effects in
$\hhvacgeon$ and $\minusvac$ sufficiently precisely to warrant an
attempt at precise terminology. For simplicity, we refer to all the
thermal properties as the Hawking effect.

\section{Scalar field theory on
$M_0$ and $M_-$}
\label{sec:flats}

In this section we discuss scalar field theory on two flat
spacetimes whose global properties mimic respectively those of the
Kruskal manifold and the $\RPthree$ geon. In subsection
\ref{subsec:flat-spaces} we construct the spacetimes as quotient
spaces of Minkowski space, and we discuss their causal and isometry
structures. In subsection
\ref{subsec:fields-on-flats} we quantize on these spacetimes a real
scalar field, using a global Minkowski-like timelike Killing vector
to define positive and negative frequencies.

\subsection{The spacetimes $M_0$ and $M_-$}
\label{subsec:flat-spaces}

Let $M$ be the $(3+1)$-dimensional Minkowski spacetime, and let
$(t,x,y,z)$ be a set of standard Minkowski coordinates on~$M$. The
metric on $M$ reads explicitly
\begin{equation}
ds^2 = - dt^2 + dx^2 + dy^2 + dz^2
\ \ .
\label{mink-metric}
\end{equation}
Let $a$ be a prescribed positive constant, and let the maps $J_0$ and
$J_-$ be defined on $M$ by
\begin{mathletters}
\label{jnoughtminus}
\begin{eqnarray}
J_0: (t,x,y,z) &\mapsto& (t,x,y,z+2a)
\ \ ,
\label{jnought}
\\
J_-: (t,x,y,z) &\mapsto& (t,-x,-y,z+a)
\ \ .
\label{jminus}
\end{eqnarray}
\end{mathletters}
$J_0$~and $J_-$ are isometries,
they preserve space orientation and time orientation,
and they act freely and properly discontinuously.
We are interested in the two quotient spaces
\begin{mathletters}
\begin{eqnarray}
&&M_0 := M/J_0
\ \ ,
\\
&&M_- := M/J_-
\ \ .
\end{eqnarray}
\end{mathletters}
By construction, $M_0$ and $M_-$ are space and time orientable flat
Lorentzian manifolds.

The universal covering space of both $M_0$ and $M_-$ is~$M$. We can
therefore construct atlases on $M_0$ and $M_-$ by using the Minkowski
coordinates $(t,x,y,z)$ as the local coordinate functions, with
suitably restricted ranges in each local chart. It will be useful to
suppress the local chart and understand $M_0$ and $M_-$ to be
coordinatized in this fashion by $(t,x,y,z)$, with the
identifications
\begin{mathletters}
\label{lor-flat-identifications}
\begin{eqnarray}
&&
(t,x,y,z) \sim (t,x,y,z+2a)
\ \ ,
\ \ \
\hbox{for $M_0$}
\ \ ,
\label{orig-idents-Mnought}
\\
&&
(t,x,y,z) \sim (t,-x,-y,z+a)
\ \ ,
\ \ \
\hbox{for $M_-$}
\ \ .
\label{orig-idents-Mminus}
\end{eqnarray}
\end{mathletters}

As $J_-^2 = J_0$, $M_0$~is a double cover of~$M_-$. $M_-$~is
therefore the quotient space of $M_0$ under the involutive isometry
${\tilde J}_-$ that $J_-$ induces on~$M_0$. In our (local)
coordinates on~$M_0$, in which the identifications
(\ref{orig-idents-Mnought}) are understood, the action of ${\tilde
J}_-$ reads as in~(\ref{jminus}).

$M_0$ and $M_-$ are static with respect to the global timelike
Killing vector~$\partial_t$. They are globally hyperbolic, and the
spatial topology of each is
$\BbbR^2 \times S^1$.\footnote{These properties remain true for
quotient spaces of $M$ with respect to arbitrary Euclidean screw
motions,
$(t,x,y,z) \mapsto
(t,x \cos\alpha - y \sin\alpha,
x \sin\alpha + y \cos\alpha,
z+b)$,
where $b\ne0$ \cite{wolf}. $J_0$~is the screw motion with $\alpha=0$
and $b=2a$, and $J_-$ is the screw motion with $\alpha=\pi$ and
$b=a$.}

$M_0$~admits seven Killing vectors. These consist of
the six Killing vectors of the
$(2+1)$-dimensional Minkowski space coordinatized by $(t,x,y)$, and
the Killing vector~$\partial_z$, which generates translations in
the compactified spacelike direction. The isometry subgroup $\BbbR^3
\times \Uone$  generated by the Killing vectors
$(\partial_t,\partial_x,\partial_y,\partial_z)$ acts on $M_0$
transitively, and $M_0$ is 
a homogeneous space \cite{wolf}.
On~$M_-$, the only Killing vectors are the time
translation Killing vector~$\partial_t$, the spacelike translation
Killing vector~$\partial_z$, and the rotational Killing vector
$x\partial_y - y\partial_x$. The isometry group of $M_-$ does not act
transitively, and $M_-$ is not a homogeneous space. One way to see
the inhomogeneity explicitly is to consider the shortest closed
geodesic in the totally geodesic hypersurface of constant~$t$.

\begin{figure}
\begin{center}
\vglue 3 cm
\leavevmode
\epsfysize=8cm
\epsfbox{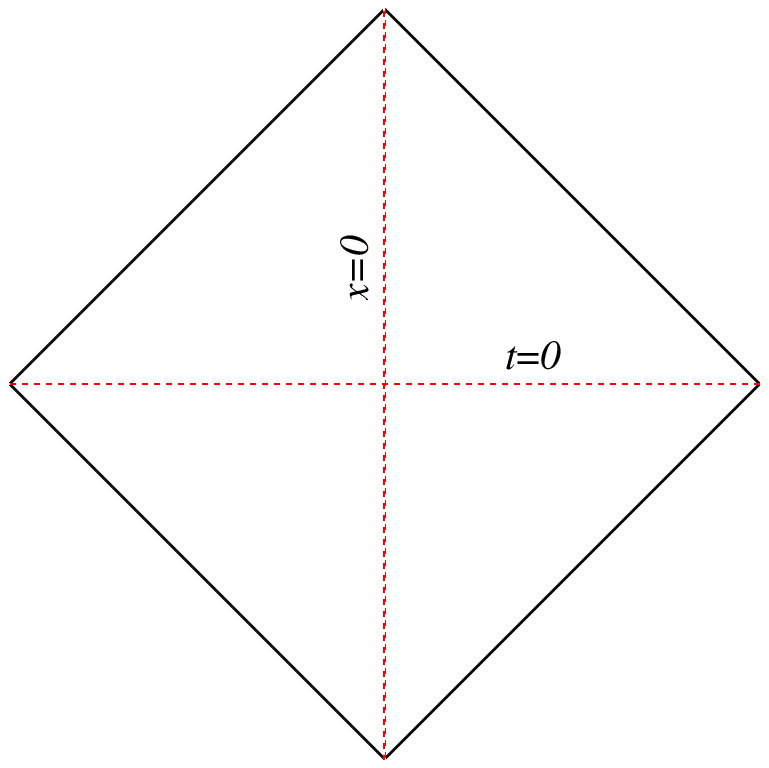}
\end{center}
\vskip 3 cm
\caption{%
A~conformal diagram of the constant $y$ and $z$ sections
of the spacetime~$M_0$.
When the diagram is understood to depict~$M_0$,
each point in the
diagram is a flat cylinder of circumference~$2a$,
coordinatized locally by $(y,z)$ with the identification
$(y,z)\sim(y,z+2a)$. Because of the suppressed
dimensions, the infinities of the diagram do not
faithfully represent the infinity structure of~$M_0$.
The involution ${\tilde J}_-$ consists
of the reflection $(t,x) \mapsto (t,-x)$
about the vertical axis, followed by the map
$(y,z) \mapsto (-y,z+a)$ on the suppressed cylinder.}
\label{fig:Mnought}
\end{figure}

It is useful to depict $M_0$ and $M_-$ in two-dimensional conformal
spacetime diagrams in which the local coordinates $y$ and $z$ are
suppressed. The diagram for~$M_0$, shown in Figure~\ref{fig:Mnought},
is that of $(1+1)$-dimensional
Minkowski spacetime. Each point in the diagram represents a flat
cylinder of circumference~$2a$, coordinatized locally by $(y,z)$ with
the identification $(y,z)\sim(y,z+2a)$. The map ${\tilde J}_-$
appears in the diagram as the reflection
$(t,x) \mapsto (t,-x)$
about the vertical axis, followed by the involution
$(y,z) \mapsto (-y,z+a)$
on the suppressed cylinder.
A~diagram that represents $M_-$ is 
obtained by taking just
the (say) right half, $x\ge0$, as shown in Figure~\ref{fig:Mminus}.
The spacetime regions depicted as $x>0$ in these two diagrams are
isometric, with each point representing a suppressed cylinder. In the
diagram for~$M_-$,
each point at $x=0$ represents
an open M\"obius strip
($\simeq\RPtwo\setminus\{\hbox{point}\}$),
with the local coordinates
$(y,z)$ identified by $(y,z)\sim(-y,z+a)$.

\subsection{Scalar field quantization with Minkowski-like vacua
on $M_0$ and~$M_-$}
\label{subsec:fields-on-flats}

We now turn to the quantum theory of a real scalar field $\phi$ with
mass $\fieldmass\ge0$ on the spacetimes $M_0$ and~$M_-$. In this
subsection we concentrate on the Minkowski-like vacua for which the
positive and negative frequencies are defined with respect to the
global timelike Killing vector~$\partial_t$.

Recall that the massive scalar field action on a general curved
spacetime is
\begin{equation}
S = - \casehalf \int \sqrt{-g} \, d^4x
\left[ g^{\mu\nu} \phi_{,\mu} \phi_{,\nu}
+ ( \fieldmass^2 + \xi R) \phi^2
\right]
\ \ ,
\label{scalar-action-gen}
\end{equation}
where $R$ is the Ricci scalar and $\xi$
is the curvature coupling constant.
On our spacetimes the Ricci scalar vanishes. In the local Minkowski
coordinates $(t,x,y,z)$, the field equation reads
\begin{equation}
\left(
- \partial_t^2 + \partial_x^2 + \partial_y^2 + \partial_z^2
- \fieldmass^2
\right)
\phi =0
\ \ .
\end{equation}
The (indefinite) inner product is
\begin{equation}
(\phi_1,\phi_2) :=
i \int_{\Sigma} \overline{\phi_1}
\,
\tensor{\partial}_t
\phi_2 \,
dx\,dy\,dz
\ \ ,
\label{inner-product}
\end{equation}
where the integration is over the constant
$t$ hypersurface~$\Sigma$. 
We denote the inner products (\ref{inner-product})
on $M_0$ and $M_-$ respectively by
${(\cdot,\cdot)}_0$ and ${(\cdot,\cdot)}_-$.

We define the positive and negative frequency solutions
to the field equation with respect to the global timelike 
Killing vector~$\partial_t$. 
It follows that
a complete orthonormal basis of positive frequency
mode functions can be built from the usual Minkowski positive
frequency mode functions as the linear combinations that are
invariant under respectively $J_0$ and~$J_-$.

On~$M_0$, a complete set of positive frequency modes
is $\left\{U_{k_x,k_y,n}\right\}$, where
\begin{equation}
U_{k_x,k_y,n} := {1 \over 4\pi  \sqrt{a\omega}}
\exp ( -i\omega t + ik_x x + i k_y y + i n\pi a^{-1} z)
\ \ ,
\end{equation}
$n\in\BbbZ$, $k_x$ and $k_y$ take all
real values, and
\begin{equation}
\omega := \sqrt{
\fieldmass^2 + k_x^2 + k_y^2 + {(n\pi/a)}^2
}
\ \ .
\label{omega-def}
\end{equation}
The orthonormality relation is
\begin{equation}
{(U_{k_x,k_y,n},U_{k'_x,k'_y,n'})}_0
= \delta_{n n'} \delta(k_x-k'_x) \delta(k_y-k'_y)
\ \ ,
\label{normalization0}
\end{equation}
with the complex conjugates satisfying a similar relation with a
minus sign, and the mixed inner products vanishing.
On~$M_-$, a complete set of positive frequency modes is
$\left\{V_{k_x,k_y,n}\right\}$, where
\begin{equation}
V_{k_x,k_y,n} := {1 \over 4\pi \sqrt{a\omega}}
\exp(-i\omega t + in \pi a^{-1} z)
\left[
\exp(ik_x x + i k_y y)
+ {(-1)}^n
\exp(-i k_x x -i k_y y)
\right]
\ \ ,
\label{V-modes}
\end{equation}
$n\in\BbbZ$, $k_x$ and $k_y$ take all
real values, and $\omega$ is as in~(\ref{omega-def}).
The orthonormality relation is
\begin{equation}
{(V_{k_x,k_y,n}, V_{k'_x,k'_y,n'})}_-
= \delta_{n n'}
\left[
\delta(k_x-k'_x) \delta(k_y-k'_y)
+
{(-1)}^n
\delta(k_x+k'_x) \delta(k_y+k'_y)
\right]
\ \ ,
\label{normalization-}
\end{equation}
with the complex conjugates again satisfying a similar relation
with a minus sign, and the mixed inner products
vanishing.\footnote{Labeling the modes (\ref{V-modes}) by the
two-dimensional momentum vector $(k_x,k_y)$ contains the
redundancy $V_{k_x,k_y,n} = {(-1)}^n V_{-k_x,-k_y,n}$. This
redundancy could be eliminated by adopting some suitable condition
(for example, $k_y>0$) that chooses a unique representative from
almost every equivalence class.}

Let $\minkvac$ denote the usual Minkowski vacuum on~$M$,
let $\noughtvac$ denote the vacuum of the set
$\left\{U_{k_x,k_y,n}\right\}$ on~$M_0$,
and let $\minusvac$ denote the vacuum of the set
$\left\{V_{k_x,k_y,n}\right\}$
on~$M_-$.
{}From the quotient space construction of $M_0$ and $M_-$ it follows
that the various two-point functions in $\noughtvac$ and $\minusvac$
can be built from the two-point functions in $\minkvac$ by the method
of images (see, for example, Ref.\ \cite{dow-ban}). If $G(x,x')$
stands for any of the usual two-point functions, this means
\begin{mathletters}
\label{G-images}
\begin{eqnarray}
G_{M_0}(x,x') &=& \sum_{n=-\infty}^\infty
G_{M}
\biglb(x,J_0^n(x') \bigrb)
\ \ ,
\label{Gnought-images}
\\
G_{M_-}(x,x') &=& \sum_{n=-\infty}^\infty
G_{M}
\biglb(x,J_-^n(x') \bigrb)
\ \ ,
\end{eqnarray}
\end{mathletters}
where $x$ and $x'$ on the right-hand side stand for points in~$M$,
while on the left-hand-side they stand for points in $M_0$ and $M_-$
in the sense of our local Minkowski coordinates.
As $J_-^2 = J_0$ and $M_-= M_0/{\tilde J}_-$, we further have
\begin{mathletters}
\begin{equation}
G_{M_-}(x,x') =
G_{M_0}
(x,x')
+
G_{M_0}
\biglb(x,{\tilde{J}}_-(x') \bigrb)
\ \ ,
\label{GMminus-image}
\end{equation}
or, more explicitly,
\begin{eqnarray}
G_{M_-}(t,x,y,z;t',x',y',z') =&&
G_{M_0} (t,x,y,z;t',x',y',z')
\nonumber
\\
&&+
G_{M_0} (t,x,y,z;t',-x',-y',z'+a)
\ \ .
\label{GMminus-image-expl}
\end{eqnarray}
\end{mathletters}

For the rest of the subsection we specialize
to a massless field, $\fieldmass=0$. The two-point functions
can then be expressed in terms of elementary functions. Consider for
concreteness the Wightman function $G^+(x,x') :=
\langle \phi(x) \phi(x') \rangle$. In~$\minkvac$, we have (see, for
example, Ref.\ \cite{birrell-davies})
\begin{equation}
G_M^+(x,x') =
{-1 \over 4 \pi^2
\left[ (t-t' -i \epsilon)^2
- (x-x')^2
- (y-y')^2
- (z-z')^2
\right]
}
\ \ ,
\label{mink-Gplus}
\end{equation}
where $\epsilon$ specifies the distributional part of $G_M^+$ in the
sense $\epsilon\to0_+$. {}From~(\ref{Gnought-images}), we find
\begin{eqnarray}
G_{M_0}^+&&(x,x')
=
{1 \over 4 \pi^2 }
\sum_{n=-\infty}^\infty
{1 \over
(z-z' + 2na)^2
+ (x-x')^2
+ (y-y')^2
- (t-t' -i \epsilon)^2
}
\nonumber
\\
=&&
{1 \over
8\pi a
\sqrt{(x-x')^2 + (y-y')^2 - (t-t' -i \epsilon)^2}
}
\nonumber
\\
&&
\times
{\sinh \left[ \pi a^{-1} \sqrt{(x-x')^2 + (y-y')^2
- (t-t' -i \epsilon)^2}
\right]
\over
\cosh \left[ \pi a^{-1} \sqrt{(x-x')^2 + (y-y')^2
- (t-t' -i \epsilon)^2}
\right] - \cos [\pi a^{-1} (z-z')]}
\ \ ,
\label{GMnought-anal}
\end{eqnarray}
where we have evaluated the sum by the calculus of residues.
$G_{M_-}^+(x,x')$ is found from (\ref{GMnought-anal})
using~(\ref{GMminus-image-expl}).

Similar calculations hold for the other two-point functions. For
example, for the Feynman propagator, one replaces $(t-t' -i
\epsilon)^2$ in (\ref{mink-Gplus}) with $(t-t')^2  -i \epsilon$,
includes an overall multiplicative factor~$-i$, and proceeds as
above.

In the Minkowski vacuum $\minkvac$ on~$M$, the renormalized
expectation value of the stress-energy tensor vanishes. As $M_0$ and
$M_-$ are flat, it is easy to find the renormalized expectation
values of the stress-energy tensor in the vacua  $\noughtvac$ and
$\minusvac$ by the point-splitting technique
\cite{birrell-davies,dow-ban}. On a Ricci-flat spacetime, the
classical stress-energy tensor computed from the action
(\ref{scalar-action-gen}) with $\fieldmass=0$ reads
\begin{equation}
T_{\mu\nu}
=
(1-2\xi) \phi_{,\mu} \phi_{,\nu}
+ (2\xi - \casehalf)
g_{\mu\nu}
g^{\rho\sigma} \phi_{,\rho} \phi_{,\sigma}
- 2 \xi \phi_{;\mu\nu} \phi
+ \casehalf \xi
g_{\mu\nu}
g^{\rho\sigma} \phi_{;\rho\sigma} \phi
\ \ .
\label{T-classical}
\end{equation}
Working in the local chart
$(t,x,y,z)$, in which $g_{\mu\nu} =
\eta_{\mu\nu} = {\rm diag}(-1,1,1,1)$, we then have, separately in
$\noughtvac$ and~$\minusvac$,
\begin{equation}
\langle T_{\mu\nu}(x) \rangle
=
\lim_{x'\to x}
{\cal D}_{\mu\nu} (x,x')
\left[
G^{(1)}(x,x') - G^{(1)}_M(x,x')
\right]
\ \ ,
\end{equation}
where $G^{(1)}(x,x') := G^+(x,x') + G^+(x',x)$ is the Hadamard
function, and the two-point differential operator ${\cal
D}_{\mu\nu} (x,x')$ reads
\begin{eqnarray}
{\cal D}_{\mu\nu} (x,x')
&=&
\case{1}{4}
(1-2\xi)
\left( \nabla_{\mu} \nabla_{\nu'}
+
\nabla_{\mu'} \nabla_{\nu}
\right)
\nonumber
\\
&&
+ \case{1}{4}
(2\xi - \casehalf)
\eta_{\mu\nu}
\left(
\eta^{\rho\sigma'} \nabla_{\rho} \nabla_{\sigma'}
+
\eta^{\rho'\sigma} \nabla_{\rho'} \nabla_{\sigma}
\right)
\nonumber
\\
&&
- \casehalf \xi
\left(
\nabla_{\mu} \nabla_{\nu}
+
\nabla_{\mu'} \nabla_{\nu'}
\right)
\nonumber
\\
&&
+ \case{1}{8} \xi
\eta_{\mu\nu}
\left(
\eta^{\rho\sigma}
\nabla_{\rho} \nabla_{\sigma}
+
\eta^{\rho'\sigma'}
\nabla_{\rho'} \nabla_{\sigma'}
\right)
\ \ .
\label{D-operator}
\end{eqnarray}
The issues of parallel transport in the operator ${\cal
D}_{\mu\nu}$ are trivial, and the renormalization has been achieved
simply by subtracting the Minkowski vacuum piece.
Using (\ref{GMminus-image-expl}) and~(\ref{GMnought-anal}),
the calculations are straightforward.
It is useful to express the final result in the
orthonormal non-coordinate frame
$\{dt,dr,\omega^{\hat\varphi},dz\}$, defined by
\begin{mathletters}
\label{polarcoords}
\begin{eqnarray}
  x &=& r \cos\varphi
  \ \ ,
  \\
  y &=& r \sin\varphi
  \ \ ,
\end{eqnarray}
\end{mathletters}
and $\omega^{\hat\varphi} := r d\varphi$.
We have
\begin{equation}
\noughtvacbra T_{\mu\nu} \noughtvac
=
{\pi^2 \over 90 {(2a)}^4} \,{\rm diag} (-1,1,1,-3)
% \ \ ,
\label{0T}
\end{equation}
and
\begin{equation}
\minusvacbra T_{\mu\nu} \minusvac
=
\noughtvacbra T_{\mu\nu} \noughtvac
+
{}^{(1)}T_{\mu\nu}
+
{}^{(2)}T_{\mu\nu}
\ \ ,
\end{equation}
where the nonvanishing components of the tensors
${}^{(1)}T_{\mu\nu}$ and
${}^{(2)}T_{\mu\nu}$
are
\begin{mathletters}
\label{1T}
\begin{eqnarray}
{}^{(1)}T_{tt}
&=&
{\pi^2 \over 4 {(2a)}^4}
\,
{1 \over s} {d \over ds}
\left( { \tanh s \over s} \right)
\ \ ,
\\
{}^{(1)}T_{zz}
&=&
{\pi^2 \over 4 {(2a)}^4}
\,
{1 \over s^2} {d \over ds}
\left( s^2 {d \over ds}  \right)
\left( { \tanh s \over s} \right)
\ \ ,
\end{eqnarray}
\end{mathletters}
\begin{mathletters}
\label{2T}
\begin{eqnarray}
{}^{(2)}T_{zz}
&=&
- {}^{(2)}T_{tt}
=
{(4\xi-1) \pi^2 \over 4 {(2a)}^4}
\,
{1 \over s} {d \over ds}  \left( s {d \over ds}  \right)
\left( { \tanh s \over s} \right)
\ \ ,
\\
{}^{(2)}T_{rr}
&=&
{(4\xi-1) \pi^2 \over 4 {(2a)}^4}
\,
{1 \over s} {d \over ds}
\left( { \tanh s \over s} \right)
\ \ ,
\\
{}^{(2)}T_{{\hat\varphi}{\hat\varphi}}
&=&
{(4\xi-1) \pi^2 \over 4 {(2a)}^4}
\,
{d^2 \over ds^2}
\left( { \tanh s \over s} \right)
\ \ ,
\end{eqnarray}
\end{mathletters}
with $s:= \pi a^{-1}\sqrt{x^2+y^2}$.

$\noughtvacbra T_{\mu\nu} \noughtvac$
and
$\minusvacbra T_{\mu\nu} \minusvac$
are conserved,
and they are clearly invariant under the
isometries of the respective spacetimes.
$\noughtvacbra T_{\mu\nu} \noughtvac$
is traceless, while
$\minusvacbra T_{\mu\nu} \minusvac$
is traceless only for conformal coupling,
$\xi=\case{1}{6}$.
At large~$r$, the difference
$\minusvacbra T_{\mu\nu} \minusvac -
\noughtvacbra T_{\mu\nu} \noughtvac$
vanishes as~$O(r^{-3})$.

\section{Uniformly accelerated observer on $M_0$ and $M_-$}
\label{sec:observer-on-flats}

In this section we consider on the spacetimes $M_0$ and $M_-$ a
uniformly accelerated observer whose world line is, in our local
Minkowski coordinates,
\begin{mathletters}
\label{observer-line}
\begin{eqnarray}
  t &=& \alpha \sinh(\tau/\alpha)
  \ \ ,
  \\
  x &=& \alpha \cosh(\tau/\alpha)
  \ \ ,
\end{eqnarray}
\end{mathletters}
with constant $y$ and~$z$. The acceleration is in the direction of
increasing~$x$, and its magnitude is~$\alpha^{-1}>0$. The parameter
$\tau$ is the observer's proper time.

In Minkowski space, it is well known that the observer
(\ref{observer-line}) sees the Minkowski vacuum $\minkvac$ as a
thermal bath at the temperature $T = {(2\pi\alpha)}^{-1}$
\cite{wald-qft,birrell-davies,takagi-magnum}.
The same conclusion is also known to hold for the vacuum $\noughtvac$
in~$M_0$ \cite{takagi-magnum}. Our purpose is to address the
experiences of the observer in the vacuum
$\minusvac$ on~$M_-$.

There are three usual ways to argue that the experiences of the
observer (\ref{observer-line}) in the Minkowski vacuum $\minkvac$ are
thermal \cite{wald-qft,birrell-davies,takagi-magnum}. First, one can
perform a Bogoliubov transformation between the Minkowski positive
frequency mode functions and the Rindler positive frequency mode
functions adapted to the accelerated observer, and in this way
exhibit the Rindler-mode content of the Minkowski vacuum.
Second, one can analyze perturbatively the response of a particle
detector that moves on the trajectory~(\ref{observer-line}). Third,
one can explore the analytic structure of the two-point functions in
the complexified time coordinate adapted to the accelerated observer,
and identify the temperature from the period in imaginary time. In
the following subsections we shall recall how these arguments
work for $\minkvac$ and~$\noughtvac$, and analyze in detail the case
of~$\minusvac$.

\subsection{Bogoliubov transformation:
non-localized Rindler modes}
\label{subsec:flat-bogo}

Consider on $M$ the Rindler wedge $|t|<x$, denoted by~$R$. We
introduce on $R$ the Rindler coordinates $(\eta,\xi,y,z)$ by
\begin{mathletters}
\label{rindler-coords}
\begin{eqnarray}
  t &=& \xi \sinh(\eta)
  \ \ ,
  \\
  x &=& \xi \cosh(\eta)
  \ \ . 
\end{eqnarray}
\end{mathletters}
These coordinates provide a global chart on~$R$,
with $\xi>0$ and $-\infty<\eta<\infty$. The metric reads
\begin{equation}
ds^2 = - \xi^2 d\eta^2 + d\xi^2 + dy^2 + dz^2
\ \ .
\label{rindler-metric}
\end{equation}
The metric (\ref{rindler-metric}) is static with respect to the
timelike Killing vector~$\partial_\eta$, which generates boosts in
the $(t,x)$ plane. In the Minkowski coordinates, $\partial_\eta = t
\partial_x + x \partial_t$.

In the Rindler coordinates, the world line (\ref{observer-line}) is
static. This suggests that the natural definition of positive and
negative frequencies for the accelerated observer is determined
by~$\partial_\eta$. One can now find the  Bogoliubov transformation
between the Rindler modes, which are defined to be positive frequency
with respect to~$\partial_\eta$, and the usual Minkowski modes, which
are positive frequency with respect to the global Killing vector
$\partial_t$ (see, for example, Ref.\ \cite{takagi-magnum}).
One finds that the Minkowski vacuum appears
as a thermal state with respect to
the Rindler modes, and the temperature seen by the observer
(\ref{observer-line}) is $T = {(2\pi\alpha)}^{-1}$. The reason why a
pure state can appear as a thermal superposition is that the Rindler
modes on $R$ do not form a complete set on~$M$: the
mixed state results from tracing over an unobserved set of
Rindler modes in the `left' wedge, $x< - |t|$.

This Bogoliubov transformation on $M$ is effectively
$(1+1)$-dimensional: the only role of the coordinates $(y,z)$ is to
contribute, through separation of variables, to the effective mass of
the $(1+1)$-dimensional modes. The transformation therefore
immediately adapts from $M$ to~$M_0$.
One concludes that the observer
(\ref{observer-line}) in $M_0$ sees the vacuum $\noughtvac$ as a
thermal state at the temperature $T = {(2\pi\alpha)}^{-1}$
\cite{takagi-magnum}.

We now turn to~$M_-$. Let ${\tilde M}_-$ denote the open region in
$M_-$ that is depicted as the `interior'
of the conformal diagram in
Figure~\ref{fig:Mminus}. {}From section \ref{sec:flats} we
recall that ${\tilde M}_-$ is isometric to the `right half'
of~$M_0$, as shown in Figure~\ref{fig:Mnought},
and it can be covered by local Minkowski coordinates
$(t,x,y,z)$ in which $x>0$ and the only identification is
$(t,x,y,z)\sim (t,x,y,z+2a)$.
We introduce on $M_-$ the Rindler wedge $R_-$ as the subset $|t|<x$
of~${\tilde M}_-$. $R_-$~is clearly isometric to the
(right-hand-side) Rindler wedge
on~$M_0$, which we denote by~$R_0$, and the observer trajectory
(\ref{observer-line}) on $M_-$ is contained in~$R_-$.

On~$R_-$, we introduce the local Rindler coordinates
$(\eta,\xi,y,z)$ by~(\ref{rindler-coords}).
The only difference from the global Rindler coordinates on
$R$ is that we now have the identification
$(\eta,\xi,y,z)\sim(\eta,\xi,y,z+2a)$.
The vector $\partial_\eta$ is a well-defined timelike
Killing vector on~$R_-$,
even though it cannot be extended
into a globally-defined Killing vector on~$M_-$.

The Rindler quantization in $R_-$ is clearly identical to that
in~$R_0$. A~complete normalized set of positive
frequency modes is $\left\{u_{\Omega,k_y,n}\right\}$, where
\cite{takagi-magnum}
\begin{equation}
u_{\Omega,k_y,n} :=
e^{i|n|\pi/2}
\sqrt{\sinh(\pi\Omega) \over 4 \pi^3 a} \,
K_{i\Omega}(\nu\xi)
\exp ( -i\Omega\eta + i k_y y + i n\pi a^{-1} z)
\ \ ,
\label{rindler-modes}
\end{equation}
$n\in\BbbZ$, $\Omega>0$, $k_y$ takes all real values,
$K_{i\Omega}$ is the modified Bessel function \cite{abra-stegun}, and
\begin{equation}
\nu := \sqrt{\fieldmass^2 + k_y^2 + {(n\pi/a)}^2}
\ \ .
\label{nu-def}
\end{equation}
% We have chosen the overall phase in (\ref{rindler-modes})
% for later convenience.
The (indefinite) inner product in~$R_-$, taken on a hypersurface of
constant~$\eta$, reads
\begin{equation}
{(\phi_1,\phi_2)}_{R_-}
:=
i \int_0^\infty {d\xi \over \xi}
\int
\,
\overline{\phi_1}
\,
\tensor{\partial}_\eta
\phi_2 \,
dy\,dz
\ \ .
\label{rindler-inner-product}
\end{equation}
The orthonormality relation is
\begin{equation}
{(u_{\Omega,k_y,n},u_{\Omega',k'_y,n'})}_{R_-}
= \delta_{n n'}
\delta(\Omega-\Omega') \delta(k_y-k'_y)
\ \ ,
\label{normalization-rindler}
\end{equation}
with the complex conjugates satisfying a similar relation with a
minus sign, and the mixed inner products vanishing.
The quantized field is expanded as
\begin{equation}
\phi =
\sum_{n=-\infty}^\infty
\int_0^\infty d\Omega
\int_{-\infty}^\infty dk_y
\left(
b_{\Omega,k_y,n} u_{\Omega,k_y,n}
+
b^\dagger_{\Omega,k_y,n} \overline{u_{\Omega,k_y,n}}
\right)
\ \ ,
\label{phi-Rminus-expansion}
\end{equation}
where the operators $b_{\Omega,k_y,n}$ and $b^\dagger_{\Omega,k_y,n}$
are the annihilation and creation operators associated with the
Rindler
mode $u_{\Omega,k_y,n}$. The Rindler vacuum $\rindvacminus$ on $R_-$
is
defined by
\begin{equation}
b_{\Omega,k_y,n} \rindvacminus =0
\ \ .
\label{rindvacminusdef}
\end{equation}

We are interested in the Rindler-mode content of the
vacuum~$\minusvac$. A direct way to proceed would be to compute
the Bogoliubov transformation between the sets
$\left\{V_{k_x,k_y,n}\right\}$ and
$\left\{u_{\Omega,k_y,n}\right\}$. However, it is easier to follow
Unruh \cite{unruh-magnum} and to build from the set
$\left\{u_{\Omega,k_y,n}\right\}$ a complete set of linear
combinations, called $W$-modes, that are bounded
analytic functions in the lower half of the complex $t$ plane.
As such modes are purely positive frequency with respect
to~$\partial_t$, their vacuum is~$\minusvac$. The Rindler-mode
content of $\minusvac$ can then be read off of the Bogoliubov
transformation that relates the set $\left\{u_{\Omega,k_y,n}\right\}$
to the $W$-modes.

In~$M_0$, the implementation of this analytic continuation argument
is well known. In the future wedge of~$M_0$,
$t>|x|$, the $W$-modes on $M_0$ are proportional to
\cite{birrell-davies-milnesec}
\begin{equation}
H^{(2)}_{i|k|}(\nu\tau)
\exp (ik\lambda + i k_y y + i n\pi a^{-1} z)
\ \ ,
\label{milne-modes-prop}
\end{equation}
where $n\in\BbbZ$,
$k$ takes all real values,
$\nu$ is given by~(\ref{nu-def}),
and $H^{(2)}_{i|k|}$ is the
Hankel function \cite{abra-stegun}.
Here $(\tau,\lambda,y,z)$ are the
Milne coordinates in the future wedge,
defined by
\begin{mathletters}
\label{milne-coords}
\begin{eqnarray}
  t &=& \tau \cosh(\lambda)
  \ \ ,
  \\
  x &=& \tau \sinh(\lambda)
  \ \ ,
\end{eqnarray}
\end{mathletters}
with $\tau>0$ and $-\infty<\lambda<\infty$.
The metric in the Milne coordinates reads
\begin{equation}
ds^2 = - d\tau^2 + \tau^2 d\lambda^2 + dy^2 + dz^2
\ \ .
\label{milne-metric}
\end{equation}
The form of the $W$-modes in the other three wedges of $M_0$ is
recovered by analytically continuing the expression
(\ref{milne-modes-prop}) across the horizons in the lower half of the
complex $t$ plane. The Bogoliubov transformation can then be read off
by comparing these $W$-modes to the Rindler modes in the right
and left Rindler wedges, $|t|<x$ and $x<-|t|$.

To develop the analogous analytic continuation
in~$M_-$, we note that the $W$-modes in the future region of $M_-$
can be built from the expressions (\ref{milne-modes-prop}) as linear
combinations that are
well defined in this region:
as the map
$J_-$ (\ref{jminus}) acts on the Milne coordinates by
$(\tau,\lambda,y,z)\mapsto(\tau,-\lambda,-y,z+a)$, the $W$-modes
are in this region proportional to
\begin{equation}
H^{(2)}_{i|k|}(\nu\tau)
\exp (i n \pi a^{-1} z)
\left[
\exp(ik\lambda + i k_y y)
+
{(-1)}^n
\exp(-ik\lambda - i k_y y)
\right]
\ \ ,
\label{minusWpropfuture}
\end{equation}
where $n\in\BbbZ$ and $\nu$ is given by~(\ref{nu-def}).
To eliminate the redundancy $(k,k_y) \to (-k,-k_y)$
in~(\ref{minusWpropfuture}), we take
$k<0$ and $-\infty<k_y<\infty$.
When analytically
continued to~$R_-$, in the lower half of the complex $t$ plane, the
expressions (\ref{minusWpropfuture}) then become proportional to
\cite{abra-stegun}
\begin{equation}
K_{i\Omega}(\nu\xi)
\exp (i n \pi a^{-1} z)
\left[
e^{\pi\Omega/2}
\exp(-i\Omega\eta + i k_y y)
+ {(-1)}^n
e^{-\pi\Omega/2}
\exp(i\Omega\eta - i k_y y)
\right]
\ \ ,
\label{minusWpropRminus}
\end{equation}
where $k$ has been renamed as~$-\Omega$, with $\Omega>0$.
Comparing (\ref{rindler-modes}) and~(\ref{minusWpropRminus}),
we see that a complete set of $W$-modes in $R_-$ is
$\left\{W_{\Omega,k_y,n}\right\}$, where
\begin{equation}
W_{\Omega,k_y,n} := {1 \over \sqrt{2 \sinh(\pi\Omega)}}
\left(
e^{\pi\Omega/2} \,
u_{\Omega,k_y,n} +
e^{-\pi\Omega/2} \,
\overline{u_{\Omega,k_y,-n}}
\right)
\ \ ,
\label{W-modes}
\end{equation}
$n\in\BbbZ$, $\Omega>0$, and $k_y$ takes all real
values.\footnote{The phase choice in
(\ref{rindler-modes}) was
made for the convenience of the phases on the
right-hand side
of~(\ref{W-modes}).}
The orthonormality relation is
\begin{equation}
{(W_{\Omega,k_y,n}, W_{\Omega',k'_y,n'})}_-
=
{(W_{\Omega,k_y,n}, W_{\Omega',k'_y,n'})}_{R_-}
= \delta_{n n'}
\delta(\Omega-\Omega') \delta(k_y-k'_y)
\ \ ,
\label{W-normalization}
\end{equation}
with the complex conjugates again satisfying a similar relation with
a minus sign, and the mixed inner products vanishing.

We can now expand the quantized field in terms of the $W$-modes as
\begin{equation}
\phi =
\sum_{n=-\infty}^\infty
\int_0^\infty d\Omega
\int_{-\infty}^\infty dk_y
\left(
d_{\Omega,k_y,n} W_{\Omega,k_y,n}
+
d^\dagger_{\Omega,k_y,n} \overline{W_{\Omega,k_y,n}}
\right)
\ \ ,
\label{phi-W-expansion}
\end{equation}
where $d_{\Omega,k_y,n}$ and $d^\dagger_{\Omega,k_y,n}$
are respectively the annihilation and creation operators
associated with the mode $W_{\Omega,k_y,n}$.
The vacuum of the $W$-modes is by
construction~$\minusvac$,
\begin{equation}
d_{\Omega,k_y,n} \minusvac =0
\ \ .
\label{minusvacdefW}
\end{equation}
Comparing the expansions (\ref{phi-Rminus-expansion})
and~(\ref{phi-W-expansion}),
and using the orthonormality relations, we find
that the Bogoliubov transformation between the annihilation and
creation operators in the two sets is
\begin{equation}
b_{\Omega,k_y,n} =
{1 \over \sqrt{2 \sinh(\pi\Omega)}}
\left(
e^{\pi\Omega/2} \,
d_{\Omega,k_y,n} +
e^{-\pi\Omega/2} \,
d^\dagger_{\Omega,k_y,-n}
\right)
\ \ ,
\label{flatbogo:b-of-d}
\end{equation}
with the inverse
\begin{equation}
d_{\Omega,k_y,n} =
{1 \over \sqrt{2 \sinh(\pi\Omega)}}
\left(
e^{\pi\Omega/2} \,
b_{\Omega,k_y,n} -
e^{-\pi\Omega/2} \,
b^\dagger_{\Omega,k_y,-n}
\right)
\ \ .
\label{flatbogo:d-of-b}
\end{equation}

We eventually wish to explore $\minusvac$ in terms of
Rindler wave packets that are localized in $\eta$ and~$y$, 
but it will be useful to postpone this to
subsection~\ref{subsec:flat-bogo-packets}, and concentrate in the
remainder of the present subsection on the content of $\minusvac$
in terms of the unlocalized Rindler modes
$\left\{u_{\Omega,k_y,n}\right\}$. We first note that
the transformation (\ref{flatbogo:d-of-b}) can be written as
\begin{equation}
d_n = \exp(-iJ) b_n \exp(iJ)
\ \ ,
\label{dbJ}
\end{equation}
where $J$ is the (formally) Hermitian operator
\begin{equation}
J := \casehalf i
\sum_{n=-\infty}^\infty r_\Omega
\left(
b_n^\dagger b_{-n}^\dagger - b_n b_{-n}
\right)
\ \ ,
\end{equation}
with $r_\Omega$ defined by
\begin{equation}
\tanh(r_\Omega) = \exp(-\pi\Omega)
\ \ .
\end{equation}
Here, 
and in the rest of this subsection, 
we suppress the labels $\Omega$ and~$k_y$. 
It follows from (\ref{rindvacminusdef}) and
(\ref{dbJ}) that $d_n \exp(-iJ) \rindvacminus =0$.
Comparing this with~(\ref{minusvacdefW}), we have
\begin{equation}
\minusvac = \exp(-iJ) \rindvacminus
\ \ .
\label{minusvac-ito-J}
\end{equation}
Expanding the exponential in (\ref{minusvac-ito-J})
and commuting the annihilation operators to
the right, we find
\begin{eqnarray}
\minusvac &=&
{1\over \sqrt{\cosh(r_\Omega)}}
\left(
\sum_{q=0}^\infty
{(2q-1)!! \, \exp(-\pi \Omega q)
\over \sqrt{(2q)!}}
|2q\rangle_0
\right)
\nonumber
\\
&&
\times
\prod_{n=1}^\infty
\left(
{1 \over \cosh(r_\Omega)}
\sum_{q=0}^\infty
\exp(-\pi \Omega q)
|q\rangle_n |q\rangle_{-n}
\right)
\ \ ,
\label{minusvac-expanded}
\end{eqnarray}
where $|q\rangle_n$ denotes the normalized state with $q$
excitations in the Rindler mode labeled by $n$ (and the suppressed
quantum numbers $\Omega$ and~$k_y$),
\begin{equation}
|q\rangle_n
:=
(q!)^{-1/2}
\left( b_n^\dagger \right)^q
\rindvacminus
\ \ .
\end{equation}
The notation in (\ref{minusvac-expanded}) is adapted to the tensor
product structure of the Hilbert space over the modes: the state
$|q\rangle_n |q\rangle_{-n}$ contains $q$ excitations both in the
mode $n$ and in the mode~$-n$. The vacuum $\minusvac$ therefore
contains Rindler excitations with $n\ne0$ in pairs whose members
only differ in the sign of~$n$.

Now, suppose that $\hat{A}$ is an operator whose support is
in~$R_-$, and suppose that $\hat{A}$ only couples to the Rindler
modes $u_{\Omega,k_y,n}$ for which $n>0$.
By~(\ref{minusvac-expanded}), the expectation value of $\hat{A}$ in
$\minusvac$ takes the form
\begin{eqnarray}
\minusvacbra
\hat{A}
\minusvac
&=&
\prod_{n=1}^\infty
\left(
[1 - \exp(-2\pi \Omega)]
\sum_{q=0}^\infty
\exp(-2q \pi \Omega)
\,
\vphantom{\rangle}_n \langle q |
\hat{A}
|q\rangle_{n}
\right)
\nonumber
\\
&=&
\hbox{tr}(\hat{A}\rho)
\ \ ,
\label{Ahat-expec-flat}
\end{eqnarray}
where
\begin{equation}
\rho =
\prod_{n=1}^\infty
\sum_{q=0}^\infty
\left[
{ \exp(-2q \pi \Omega)
\over
\sum_{m=0}^\infty \exp(-2m \pi \Omega)
}
\right]
|q\rangle_{n}
\,
\vphantom{\rangle}_n \langle q |
\ \ .
\label{rho-flat}
\end{equation}
The operator $\rho$ has the form of a thermal density matrix.
Specializing to an $\hat{A}$ that is concentrated on the
accelerated world line~(\ref{observer-line}), we infer from
equations (\ref{Ahat-expec-flat}) and~(\ref{rho-flat}), and the
redshift in the metric~(\ref{rindler-coords}), that the accelerated
observer sees the operator $\hat{A}$ as coupling to a thermal bath
at the temperature
$T = {(2\pi\alpha)}^{-1}$ \cite{birrell-davies,takagi-magnum}.
A~similar result clearly holds when $\hat{A}$ is replaced by any
operator that does not couple to the modes with $n=0$ and, for each
triplet $(\Omega,k_y,n)$ with $n\ne0$, only couples to one of the
modes $u_{\Omega,k_y,n}$ and $u_{\Omega,k_y,-n}$. For operators that
do not satisfy this property, on the other hand, the experiences of
the accelerated observer are not thermal.

It is instructive to contrast these results on $M_-$ to their
well-known counterparts on $M_0$
\cite{birrell-davies,takagi-magnum}. On $M_0$ there are  two sets
of Rindler modes, one set for the 
right-hand-side
Rindler wedge and the other for the 
left-hand-side
Rindler wedge. There are also twice as many $W$-modes
as on~$M_-$, owing to the fact the modes (\ref{milne-modes-prop})
are distinct for positive and negative values of~$k$. The
counterpart of (\ref{flatbogo:b-of-d}) consists of the two equations
\begin{mathletters}
\label{flatbogo:Mnought}
\begin{eqnarray}
&&b^{(1)}_{\Omega,k_y,n} =
{1 \over \sqrt{2 \sinh(\pi\Omega)}}
\left[
e^{\pi\Omega/2} \,
d^{(1)}_{\Omega,k_y,n} +
e^{-\pi\Omega/2} \,
{\left(d^{(2)}_{\Omega,k_y,n}\right)}^\dagger
\right]
\ \ ,
\\
&&b^{(2)}_{\Omega,k_y,n} =
{1 \over \sqrt{2 \sinh(\pi\Omega)}}
\left[
e^{\pi\Omega/2} \,
d^{(2)}_{\Omega,k_y,n} +
e^{-\pi\Omega/2} \,
{\left(d^{(1)}_{\Omega,k_y,n}\right)}^\dagger
\right]
\ \ ,
\end{eqnarray}
\end{mathletters}
where the superscript on the $b$'s indicates the Rindler wedge,
and the superscript on the $d$'s serves as an additional label on
the $W$-modes in a way whose details are not relevant here. The
counterpart of (\ref{minusvac-expanded}) on $M_0$ therefore reads
\begin{equation}
\noughtvac =
\prod_{n=-\infty}^\infty
\left(
{1 \over \cosh(r_\Omega)}
\sum_{q=0}^\infty
\exp(- \pi \Omega q)
|q\rangle_n^{(1)} |q\rangle_n^{(2)}
\right)
\ \ ,
\label{noughtvac-expanded}
\end{equation}
where the superscripts again indicate the Rindler wedge. For any
operator $\hat{A}^{(1)}$ on $M_0$ whose support is in the wedge
labeled by the superscript~$(1)$, we obtain
\begin{eqnarray}
\noughtvacbra
\hat{A}^{(1)}
\noughtvac
&=&
\prod_{n=-\infty}^\infty
\left(
[1 - \exp(-2\pi \Omega)]
\sum_{q=0}^\infty
\exp(-2q \pi \Omega)
\,
\vphantom{\rangle}_{\;\;n}^{(1)} \langle q |
\hat{A}^{(1)}
|q\rangle_{n}^{(1)}
\right)
\nonumber
\\
&=&
\hbox{tr}(\hat{A}^{(1)}\rho^{(1)})
\ \ ,
\label{Ahat-expec-flat-Mnought}
\end{eqnarray}
where
\begin{equation}
\rho^{(1)} =
\prod_{n=-\infty}^\infty
\sum_{q=0}^\infty
\left[
{ \exp(-2q \pi \Omega)
\over
\sum_{m=0}^\infty \exp(-2m \pi \Omega)
}
\right]
|q\rangle_{n}^{(1)}
\,
\vphantom{\rangle}_{\;\;n}^{(1)} \langle q |
\ \ .
\label{rho-flat-Mnought}
\end{equation}
$\rho^{(1)}$ has the form of a thermal density matrix. On~$M_0$,
equations (\ref{Ahat-expec-flat-Mnought}) and
(\ref{rho-flat-Mnought}) hold now for any operator whose support is
on the accelerated world line~(\ref{observer-line}), regardless how
this operator couples to the various Rindler modes. One infers that
the accelerated observer on $M_0$ sees a thermal bath at the
temperature $T = {(2\pi\alpha)}^{-1}$
\cite{birrell-davies,takagi-magnum}, no matter what (local)
operators the observer may employ to probe~$\noughtvac$.

Finally, let us consider number operator
expectation values. Using respectively (\ref{flatbogo:b-of-d})
and~(\ref{flatbogo:Mnought}), we find
\begin{eqnarray}
\minusvacbra
b^\dagger_{\Omega,k_y,n}
b_{\Omega',k_y',n'}
\minusvac
\nonumber
&=&
\noughtvacbra
\left( b^{(1)}_{\Omega,k_y,n}\right)^\dagger
b^{(1)}_{\Omega',k_y',n'}
\noughtvac
\nonumber
\\
&=&
{\left( e^{2\pi\Omega} - 1 \right)}^{-1}
\delta_{n n'}
\delta(\Omega-\Omega') \delta(k_y-k'_y)
\ \ .
\label{pre-number-planckian}
\end{eqnarray}
Setting the primed and unprimed indices equal in
(\ref{pre-number-planckian}) shows that the number operator
expectation value of a given Rindler mode is divergent both in
$\noughtvac$ and~$\minusvac$. This divergence arises from the
delta-function-normalization of our mode functions, and it
disappears when one introduces finitely-normalized Rindler wave
packets \cite{hawkingCMP,takagi-magnum}. What can be immediately
seen from  (\ref{pre-number-planckian}) is, however, that the
number operator expectation values are identical in $\noughtvac$
and~$\minusvac$, even after introducing normalized wave packets. We
shall return to this point in the next subsection.

\subsection{Bogoliubov transformation:
Rindler wave packets}
\label{subsec:flat-bogo-packets}

In subsection \ref{subsec:flat-bogo} we explored $\minusvac$ in
terms of Rindler modes that are unlocalized in $\eta$ and~$y$.
While translations in $\eta$ and $y$ are isometries of the Rindler
wedge~$R_-$, the restriction of $\minusvac$ to $R_-$ is not
invariant under these isometries, as is evident from the isometry
structure of the two-point functions in~$\minusvac$. This
suggests that more information about $\minusvac$ can be unraveled
by using Rindler wave packets that are localized in $\eta$
and~$y$. In this subsection we consider such modes.

For concreteness, we form wave packets following closely Refs.\ 
\cite{hawkingCMP,takagi-magnum}. As a preliminary, let $\epsilon_1>0$,
and define the functions $f_{ml}: \BbbR \to \BbbC$, with
$m,l\in\BbbZ$, by
\begin{mathletters}
\label{fh-defs}
\begin{equation}
f_{ml}(k) :=
\cases{
\epsilon_1^{-1/2}
\exp( -2\pi i \epsilon_1^{-1} l k)
\ ,
\ \
&for $m\epsilon_1 < k < (m+1) \epsilon_1$
\ ,
\cr
0
\ ,
\ \
&otherwise
\ .
\cr}
\label{ffd}
\end{equation}
Similarly, let $\epsilon_2>0$, and define the functions
$h_{\rho\sigma}: \BbbR_+ \to \BbbC$, with $\rho,\sigma\in\BbbZ$ and
$\rho\ge0$, by
\begin{equation}
h_{\rho\sigma}(\Omega) :=
\cases{
\epsilon_2^{-1/2}
\exp( -2\pi i \epsilon_2^{-1} \sigma \Omega)
\ ,
\ \
&for $\rho\epsilon_2 < \Omega < (\rho+1) \epsilon_2$
\ ,
\cr
0
\ ,
\ \
&otherwise
\ .
\cr}
\label{hfd}
\end{equation}
\end{mathletters}
These functions satisfy the orthonormality and completeness
relations \cite{takagi-magnum}
\begin{mathletters}
\begin{eqnarray}
&&
\int_{-\infty}^\infty
dk \,
\overline{ f_{ml}(k) }
f_{m'l'}(k)
=
\delta_{m m'} \delta_{l l'}
\ \ ,
\\
&&
\int_0^\infty
d\Omega \,
\overline{ h_{\rho\sigma}(\Omega) }
h_{\rho'\sigma'}(\Omega)
=
\delta_{\rho \rho'} \delta_{\sigma \sigma'}
\ \ ,
\\
&&
\sum_{m l}
\overline{ f_{ml}(k) }
f_{ml}(k')
=
\delta(k-k')
\ \ ,
\\
&&
\sum_{\rho \sigma}
\overline{ h_{\rho\sigma}(\Omega) }
h_{\rho\sigma}(\Omega')
=
\delta(\Omega-\Omega')
\ \
\end{eqnarray}
\end{mathletters}

We define the Rindler wave packets
$\left\{u_{\rho\sigma m l n}\right\}$ by
\begin{mathletters}
\label{uloc-def-pair}
\begin{equation}
u_{\rho\sigma m l n}
:=
\int_0^\infty d\Omega
\int_{-\infty}^\infty dk_y
\,
h_{\rho\sigma}(\Omega)
f_{ml}(k_y)
\,
u_{\Omega,k_y,n}
\ \ .
\label{uloc-def}
\end{equation}
It is easily verified that the set
$\left\{u_{\rho\sigma m l n}\right\}$
is complete and orthonormal in the Klein-Gordon inner product, and
that the inverse of (\ref{uloc-def}) reads
\begin{equation}
u_{\Omega,k_y,n}
=
\sum_{\rho\sigma m l}
\overline{ h_{\rho\sigma}(\Omega) f_{ml}(k_y) }
\,
u_{\rho\sigma m l n}
\ \ .
\label{uloc-def-inverse}
\end{equation}
\end{mathletters}
The annihilation and creation operators associated with the mode
$u_{\rho\sigma m l n}$ are denoted by $b_{\rho\sigma m l n}$ and
$b^\dagger_{\rho\sigma m l n}$. {}From~(\ref{uloc-def-pair}), we
then have the relations
\begin{mathletters}
\label{bloc-def-pair}
\begin{eqnarray}
b_{\rho\sigma m l n}
&=&
\int_0^\infty
d\Omega
\int_{-\infty}^\infty
d k_y \,
\overline{ h_{\rho\sigma}(\Omega) f_{ml}(k_y) }
\,
b_{\Omega,k_y,n}
\ \ ,
\label{bloc-def}
\\
b_{\Omega,k_y,n}
&=&
\sum_{\rho\sigma m l}
h_{\rho\sigma}(\Omega) f_{ml}(k_y)
\,
b_{\rho\sigma m l n}
\ \ .
\label{bloc-def-inverse}
\end{eqnarray}
\end{mathletters}

It is clear from the definition that the mode $u_{\rho\sigma m l
n}$ is localized in $k_y$ around the value
$(m+\casehalf)\epsilon_1$ with width~$\epsilon_1$, and in $\Omega$
around the value $\Omega_\rho := (\rho+\casehalf)\epsilon_2$ with
width~$\epsilon_2$. What is important for us is that the mode is
approximately localized also in $\eta$ and~$y$. When the
$k_y$-dependence of the modified Bessel function
$K_{i\Omega}(\nu\xi)$ in (\ref{rindler-modes}) [via~$\nu$,
(\ref{nu-def})] can be 
ignored\footnote{For example, for $\epsilon_1 \ll  \nu$ or 
$\Omega  \gg \nu$.}
in the integral~(\ref{uloc-def}),
one sees as in Ref.\ \cite{takagi-magnum} that $u_{\rho\sigma m l
n}$ is approximately localized in $y$ around $y_l := 2\pi
\epsilon_1^{-1} l$ with width~$2\pi\epsilon_1^{-1}$. Similarly,
when $\Omega$ is large enough that
$\nu\xi\ll\Omega$, $K_{i\Omega}(\nu\xi)$
is proportional to a linear combination of two terms whose
$\xi$-dependence is $\xi^{\pm i\Omega}$, and it follows as in Ref.\
\cite{takagi-magnum} that, at fixed $\xi$,
$u_{\rho\sigma m l n}$ is approximately
localized in $\eta$ around two peaks, situated at $\eta = - 2\pi
\epsilon_2^{-1} \sigma \pm \ln\xi$, and each having
width~$2\pi\epsilon_2^{-1}$. We can therefore understand
$u_{\rho\sigma m l n}$ to be localized at large positive (negative)
values of $y$ for large positive (negative)~$l$, and, for
given~$\xi$, at large positive (negative) values of $\eta$ for
large negative (positive)~$\sigma$. While this leaves the sense of
the localization somewhat imprecise, especially regarding the
uniformity of the localization with respect to $\xi$ and the
various parameters of the modes, this discussion will nevertheless
be sufficient for obtaining qualitative results about the vacuum
$\minusvac$ in the limits of interest.
We will elaborate further on the technical details below.

In order to write $\minusvac$ in terms of the operators
$b^\dagger_{\rho\sigma m l n}$ acting on~$\rindvacminus$, 
we define the $W$-packets $\left\{W_{\rho\sigma m l
n}\right\}$ by a formula analogous to~(\ref{uloc-def}), with
$u_{\Omega,k_y,n}$ replaced by~$W_{\Omega,k_y,n}$. Denoting by
$d_{\rho\sigma m l n}$ and $d^\dagger_{\rho\sigma m l n}$
the annihilation and creation operators associated with the mode
$W_{\rho\sigma m l n}$, we have for
$d_{\rho\sigma m l n}$ and $d_{\Omega,k_y,n}$ a pair of relations
analogous to~(\ref{bloc-def-pair}).
{}From~(\ref{flatbogo:b-of-d}), we then obtain
\begin{equation}
b_{\rho\sigma m l n} =
\epsilon_2^{-1}
\sum_{\sigma'}
\int_{\rho\epsilon_2}
^{(\rho+1)\epsilon_2}
d\Omega \,
\frac{
\exp[ 2\pi i \epsilon_2^{-1}
(\sigma - \sigma') \Omega]
}
{\sqrt{2 \sinh(\pi\Omega)}}
\left(
e^{\pi\Omega/2} \,
d_{\rho \sigma' m l n} +
e^{-\pi\Omega/2} \,
d^\dagger_{\rho (-\sigma') m (-l) (-n)}
\right)
\ .
\label{flatbogopac:b-of-d}
\end{equation}
We now assume $\epsilon_2\ll1$. Equation (\ref{flatbogopac:b-of-d})
can then be approximated by
\begin{equation}
b_{\rho\sigma m l n} \approx
\frac{1}
{\sqrt{2 \sinh(\pi\Omega_\rho)}}
\left(
e^{\pi\Omega_\rho/2} \,
d_{\rho \sigma m l n} +
e^{-\pi\Omega_\rho/2} \,
d^\dagger_{\rho (-\sigma) m (-l) (-n)}
\right)
\ \ .
\label{flatbogopacapp:b-of-d}
\end{equation}
Comparing (\ref{flatbogopacapp:b-of-d}) to
(\ref{flatbogo:b-of-d}) and proceeding as in
subsection~\ref{subsec:flat-bogo}, we find
\begin{eqnarray}
\minusvac 
\approx
\prod_{\rho m}
&&
\left[
\vphantom{\mathop{{\prod}'}_{\![\sigma l n]}}
{1\over \sqrt{\cosh(r_{\Omega_\rho})}}
\left(
\sum_{q=0}^\infty
{(2q-1)!! \, \exp(-\pi \Omega_\rho q)
\over \sqrt{(2q)!}}
|2q\rangle_{\rho 0 m 0 0}
\right)
\right.
\nonumber
\\
&&
\quad 
\left.
\times
\mathop{{\prod}'}_{\![\sigma l n]}
\left(
{1 \over \cosh(r_{\Omega_\rho})}
\sum_{q=0}^\infty
\exp(-\pi \Omega_\rho q)
|q\rangle_{\rho \sigma m l n}
|q\rangle_{\rho (-\sigma) m (-l) (-n)}
\right)
\right]
\ \ ,
\label{minusvac-expanded-pac}
\end{eqnarray}
where $|q\rangle_{\rho \sigma m l n}$ denotes the normalized state
with $q$ excitations in the mode $u_{\rho \sigma m l n}$,
\begin{equation}
|q\rangle_{\rho \sigma m l n}
:=
(q!)^{-1/2}
\left( b_{\rho \sigma m l n}^\dagger \right)^q
\rindvacminus
\ \ .
\end{equation}
The primed product $\mathop{{\prod}'}_{\![\sigma l n]}$ is over
all equivalence classes $[\sigma l n]$ of triples under the
identification $(\sigma, l, n) \sim (-\sigma, -l, -n)$, except the
equivalence 
class $[000]$. 

Comparing (\ref{minusvac-expanded-pac})
to~(\ref{minusvac-expanded}), we see that the expectation values in
$\minusvac$ are thermal for any operator that does not couple to
the modes with $\sigma=l=n=0$, and, for fixed $\rho$ and~$m$, only
couples to one member of each equivalence class $[\sigma l n] \ne
[000]$. Because of the mode localization properties discussed
above, the accelerated observer (\ref{observer-line}) at early
(late) times only couples to modes with large positive (negative)
values of~$\sigma$, and thus sees $\minusvac$ as a thermal state in
the temperature $T = {(2\pi\alpha)}^{-1}$. Similarly, if the world
line of the observer is located at a large positive (negative) value
of~$y$, the observer only couples to modes with large positive
(negative) values of~$l$, and sees $\minusvac$ as thermal in the
same temperature. In these limits, the observer thus cannot
distinguish between the vacua $\minusvac$ and~$\noughtvac$.

We note that,
in these limits, the observer is in a region of 
spacetime where
$\minusvacbra T_{\mu\nu} \minusvac$
and
$\noughtvacbra T_{\mu\nu} \noughtvac$
for a massless field agree, as seen in
subsection~\ref{subsec:fields-on-flats}.
The same property seems likely to hold also for the
stress-energy tensor of a massive field.

The correlations exhibited in (\ref{minusvac-expanded-pac})
should not be surprising. To see this, consider the analogue of
(\ref{minusvac-expanded-pac}) for the vacuum
$\noughtvac$ on $M_0$ \cite{takagi-magnum}.
{}From the invariance of $\noughtvac$ under the
isometries of $M_0$ it follows that a right-hand-side Rindler
packet localized at early (late) right-hand-side Rindler times is
correlated with a left-hand-side Rindler packet localized at late
(early) left-hand-side Rindler times, and that a right-hand-side
Rindler packet localized at large positive (negative) $y$ is
correlated with a left-hand-side Rindler packet localized at large
positive (negative)~$y$. As the map ${\tilde J}_-$ on $M_0$ takes
late (early) right-hand-side Rindler times to late (early)
left-hand-side
Rindler times, and inverts the sign of~$y$, one
expects that in~$\minusvac$, a Rindler-packet localized at early
Rindler times should be correlated with a packet localized at late
Rindler times, and a packet localized at large positive $y$ should
be correlated with a packet localized at large negative~$y$. This
is exactly the structure displayed
by~(\ref{minusvac-expanded-pac}).

Finally, consider the number operator expectation value of the mode
$u_{\rho\sigma m l n}$ in~$\minusvac$. Using
(\ref{pre-number-planckian}) and~(\ref{bloc-def}), we obtain
\begin{eqnarray}
N_{\rho\sigma m l n}
&:=&
\minusvacbra
b^\dagger_{\rho\sigma m l n}
b_{\rho\sigma m l n}
\minusvac
\nonumber
\\
&=&
\epsilon_2^{-1}
\int_{\rho\epsilon_2}
^{(\rho+1)\epsilon_2}
d\Omega \,
{\left( e^{2\pi\Omega} - 1 \right)}^{-1}
\ \ .
\label{loc-number-planckian}
\end{eqnarray}
As $\epsilon_2\ll1$, (\ref{loc-number-planckian}) yields
\begin{equation}
N_{\rho\sigma m l n}
\approx
{\left( e^{2\pi\Omega_\rho} - 1 \right)}^{-1}
\ \ .
\label{app-loc-number-planckian}
\end{equation}
The spectrum for $N_{\rho\sigma m l n} $ is thus Planckian and,
taking into account the redshift to the local frequency seen by the
accelerated observer, corresponds to the temperature $T =
{(2\pi\alpha)}^{-1}$. The result
(\ref{loc-number-planckian}) is precisely the same as in
the vacuum $\minusvac$ on $M_0$ \cite{takagi-magnum}, as noted at
the end of subsection~\ref{subsec:flat-bogo}. The number operator
expectation value thus contains no information about the
noninvariance of $\minusvac$ under translations in $\eta$ and~$y$.

In the above analysis, we have so far justified 
the localization arguments in $\eta$ only for modes with 
$\Omega \gg \nu \xi$.  
These are the modes where the radial momentum is large
enough that the mode behaves relativistically out to this location 
({\em i.e.,} the effective mass $\nu$ for radial propagation 
is irrelevant out to~$\xi$).  
As a result, the radial propagation is that of a 
$(1+1)$-dimensional free
scalar field, with minimal spreading and dispersion.
In fact, even in this case, we did not discuss the uniformity of our
approximations, and it turns out that the localized modes defined by
(\ref{fh-defs}) are somewhat too broad to be of use in
a rigorous analysis.  The point here is that, due to the sharp corners
of the step functions in~(\ref{fh-defs}), 
the modes $\left\{u_{\rho\sigma m l n}\right\}$
have long tails that decay only as $\eta^{-1}$ or $y^{-1}$, too 
slowly for convergence of certain integrals.  However, this can be
handled in the usual ways, for example by 
wavelet techniques \cite{wavelets}.

Although it is more complicated to 
discuss in detail, the lower energy
modes (where $\Omega \gg \nu \xi$ does not hold)  
are also well localized
in~$\eta$.  For the following discussion, let us ignore the
$y$ and $z$ directions except as they contribute to the effective
mass for propagation in the $(\eta,\xi)$-plane.  Our discussion will
make use of the fact [see equation~(\ref{hfd})]
that replacing the index $\sigma$
on the mode $u_{ \rho \sigma m l n}$ with $\sigma + \tau$ is 
equivalent to a translation of the mode under $\eta \rightarrow \eta
+  2 \pi \tau/ \epsilon_2$.  Thus, if any mode is localized in $\eta$ 
(for fixed~$\xi$), the localization 
is determined by the value of~$\sigma$. 
In particular, for large positive $\sigma$  
($\sigma \gg \epsilon_2$), the mode
will be localized at $\eta \gg 1$, while for large negative
$\sigma$ ($\sigma \ll - \epsilon_2$) the mode will be localized at 
$\eta  \ll -1$.  Thus, we need only show that at fixed $\xi$
the lower energy modes
decay rapidly as $\eta \rightarrow \pm \infty$ 
in order to show that operators at late
times couple only to modes with large positive~$\sigma$.

Consider those modes with energy $\Omega \sim \nu \xi$.
We will address such modes through the equivalence principle.
{}From this perspective, the modes with $\Omega \ll \nu \xi$ are
those modes that do not have sufficient energy to climb to the height
$\xi$ in the effective
gravitational field.  Similarly, the modes with 
$\Omega \gg \nu \xi$ have so much energy that, not only can they
climb to the height $\xi$, but that they remain relativistic in this
region and so continue to propagate with minimal dispersion.   
Thus, we see that the modes with $\Omega \sim \nu \xi$ are those modes
that, while they have sufficient energy to reach the vicinity of~$\xi$, 
propagate nonrelativistically through this region.  Thus, 
we may describe them as the wave functions 
of nonrelativistic particles in a gravitational field.

For large times, 
it is reasonable to model the corresponding 
wave packets by ignoring the
effect of the gravitational field on the 
dispersion of the packet and only
including this field through its effects 
on the center of the wave packet.
That is, we model such a wave packet as the wave packet of a free 
nonrelativistic particle
for which, instead of following a constant velocity trajectory, the
center of the packet accelerates downward as described by the field.
Such an estimate of the large time behavior at fixed position
gives an exponential decay of the 
wave function as the packet `falls
down the gravitational well.'  Thus, we conclude that 
the mode decays exponentially with the proper time 
(proportional to~$\eta$) at any location $\xi$.  
It follows that modes with
$\Omega \ll \nu \xi$ should also have at least exponential decay in 
$\eta$ at fixed~$\xi$, since they do not even have enough energy to
classically reach the height~$\xi$.  Thus, even for modes
that do not satisfy $\Omega \gg \nu \xi$, we conclude that operators
at large positive $\eta$ couple only to modes with large 
positive~$\sigma$, and so view $\minusvac$ as a thermal bath.

\subsection{Particle detector}
\label{subsec:flat-detector}

In this subsection we consider on the spacetimes $M_0$ and $M_-$
a monopole detector whose world line is given
by~(\ref{observer-line}). The detector is turned on and off in a
way to be explained below, and the detector ground state energy is
normalized to~$0$. The field is taken to be massless.

In first order perturbation theory, the probability for the detector
becoming excited is
\cite{unruh-magnum,dewitt-CC,birrell-davies,takagi-magnum}
\begin{equation}
c^2 \sum_{E>0}
{\left|
\vphantom{|^A_A}
\langle\!\langle E | \bbox{m}(0) | 0 \rangle\!\rangle \right|}^2
{\cal{F}}(E)
\ \ ,
\label{exci-prob}
\end{equation}
where $c$ is the coupling constant,
$\bbox{m}(\tau)$ is the detector's
monopole moment operator,
$| 0 \rangle\!\rangle$ is the ground state of the detector,
the sum is over all the excited states
$| E \rangle\!\rangle$ of the detector, and
\begin{equation}
{\cal{F}}(E)
:=
\int d\tau \int d\tau' \,
e^{-iE(\tau-\tau')}
G^+ \biglb( x(\tau), x(\tau') \bigrb)
\ \ .
\label{resp-function}
\end{equation}
Here $G^+(x,x')$ stands on $M_0$ for the Wightman function
$G_{M_0}^+(x,x')
:= \noughtvacbra \phi(x) \phi(x') \noughtvac$,
and on $M_-$ for the Wightman function
$G_{M_-}^+(x,x')
:= \minusvacbra \phi(x) \phi(x') \minusvac$.
The detector response function ${\cal{F}}(E)$ contains the
information about the environment (by definition, `particles')
seen by the detector,
while the remaining factor in (\ref{exci-prob}) contains the
information about the sensitivity by the detector.

Consider first the vacuum $\noughtvac$ on~$M_0$. The Wightman
function $G_{M_0}^+(x,x')$ is by construction invariant under the
boosts generated by the Killing vector $x \partial_t + t \partial_x$.
We therefore have
$G_{M_0}^+ \biglb( x(\tau), x(\tau') \bigrb) =
G_{M_0}^+ \biglb( x(\tau-\tau'), x(0) \bigrb)$.
This implies that the excitation probability per unit
proper time is constant. In particular, if the detector is turned
on at the infinite past and off at the infinite future, each
integral in (\ref{resp-function}) has a fully infinite range, and
the total probability (\ref{exci-prob}) is either divergent or zero.
A~more meaningful quantity in this instance is the excitation
probability per unit proper time, given by the counterpart of
(\ref{exci-prob}) with ${\cal{F}}(E)$ replaced by
\begin{equation}
{\tilde{\cal{F}}}(E)
:=
\int d\tau \,
e^{-iE\tau}
G^+ \biglb( x(\tau), x(0) \bigrb)
\ \ .
\label{resp-function-norm}
\end{equation}
Inserting the trajectory (\ref{observer-line})
into the image sum expression in~(\ref{GMnought-anal}), 
we find 
\begin{equation}
G_{M_0}^+ \biglb( x(\tau), x(0) \bigrb)
=
{- 1 \over 16 \pi^2 \alpha^2}
\sum_{n=-\infty}^\infty
{1 \over
\sinh^2 [(\tau-i\epsilon)/(2\alpha)] - n^2 a^2 \alpha^{-2}
}
\ \ .
\label{Gmnought-accsum}
\end{equation}
The contributions to 
(\ref{resp-function-norm})
from each term in (\ref{Gmnought-accsum})
can then be evaluated by contour 
intergrals, with the result 
\begin{equation}
{\tilde{\cal{F}}}_{M_0}(E)
=
{E \over 2\pi
\left( e^{2\pi\alpha E} -1 \right)
}
\left(
1 + \sum_{n=1}^\infty
{\sin[2\alpha E \arcsinh(na/\alpha)]
\over
n a E\sqrt{1 + n^2 a^2 \alpha^{-2}}
}
\right)
\ \ .
\label{Mnought-Ftildef}
\end{equation}
${\tilde{\cal{F}}}_{M_0}(E)$ clearly satisfies the KMS condition
\begin{equation}
{\tilde{\cal{F}}}_{M_0}(E)
= e^{-2\pi\alpha E}
{\tilde{\cal{F}}}_{M_0}(-E)
\ \ ,
\end{equation}
which is characteristic of a thermal response at the temperature $T =
{(2\pi\alpha)}^{-1}$ \cite{takagi-magnum}. In the limit
$a\to\infty$, only the first term in (\ref{Mnought-Ftildef})
survives, and ${\tilde{\cal{F}}}_{M_0}(E)$ correctly reduces to
${\tilde{\cal{F}}}_{M}(E)$ \cite{birrell-davies}.

Consider then the vacuum $\minusvac$ on~$M_-$.
{}From (\ref{GMminus-image}) and (\ref{GMnought-anal}) we obtain
\begin{equation}
G_{M_-}^+ \biglb( x(\tau),
x(\tau') \bigrb)
=
G_{M_0}^+ \biglb( x(\tau),
x(\tau') \bigrb)
+
\Delta G^+(\tau,\tau')
\ \ ,
\end{equation}
where
\begin{equation}
\Delta G^+(\tau,\tau')
=
{\tanh \!
\left\{
(\pi/a)
\sqrt{ \alpha^2 \cosh^2 [(\tau+\tau')/(2\alpha)] + y_0^2}
\right\}
\over
16 \pi a
\sqrt{ \alpha^2 \cosh^2 [(\tau+\tau')/(2\alpha)] + y_0^2}
}
\ \ ,
\label{Delta-Gplus}
\end{equation}
and $y_0$ is the value of the coordinate
$y$ on the detector trajectory.
If the detector is turned on
at the infinite past and off at the infinite future,
the contribution from $\Delta G^+(\tau,\tau')$
to ${\cal{F}}_{M_-}(E)$
is equal to a finite number times~$\delta(E)$.
This implies that $\Delta G^+(\tau,\tau')$ does not contribute to
the (divergent) total excitation probability~(\ref{exci-prob}).
However, the excitation probability per unit proper time is now not
a constant along the trajectory, since $\Delta G^+(\tau,\tau')$
depends on $\tau$ and $\tau'$ through the sum $\tau+\tau'$. 
The vacua $\minusvac$ and $\noughtvac$ 
appear therefore distinct to particle
detectors that only operate for some finite duration, and it is not
obvious whether the response of such detectors in $\minusvac$ can be
regarded as thermal.
Nevertheless, the suppression of $\Delta G^+(\tau,\tau')$ at large
$|\tau+\tau'|$ shows that the responses in $\minusvac$ and
$\noughtvac$ are asymptotically identical for a detector that only
operates in the asymptotic past or future. Similarly, the
suppression of $\Delta G^+(\tau,\tau')$ at large $|y_0|$ shows that
the responses in $\minusvac$ and $\noughtvac$ become asymptotically
identical for a detector whose trajectory lies at asymptotically
large values of~$|y|$, uniformly for all proper times along the
trajectory. The detector in $\minusvac$ therefore responds
thermally, at the temperature $T = {(2\pi\alpha)}^{-1}$, in the
limit of early and late proper times for a prescribed~$y_0$, and
for all proper times in the limit of large~$|y_0|$.
These are precisely the limits in which we deduced the experiences
along the accelerated world line to be thermal from the Bogoliubov
transformation in subsection~\ref{subsec:flat-bogo-packets}.

\subsection{Riemannian section and the periodicity of
Riemannian Rindler time}
\label{subsec:flat-imagtime}

In this subsection we consider the  analytic properties of the
Feynman Green functions  in the complexified Rindler time coordinate.
We begin by discussing the relevant Riemannian sections of the
complexified spacetimes.

As~$M$, $M_0$, and $M_-$ are static,
they can be regarded as Lorentzian
sections of complexified flat spacetimes that
also admit Riemannian sections.
In terms of the (local) coordinates $(t,x,y,z)$,
the Riemannian sections of interest arise by writing
$t = -i\ttilde$, letting the `Riemannian time' coordinate
$\ttilde$ take all real values, and keeping $x$, $y$, and $z$
real.\footnote{The Lorentzian and Riemannian
sections could be defined as the
sets stabilized by suitable antiholomorphic involutions on the
complexified spacetimes. We shall rely on this formalism
with the black hole spacetimes in section
\ref{sec:kruskal-rpthree-defs}
\cite{gibb-holo,chamb-gibb}.}
We denote the resulting flat
Riemannian manifolds by respectively
$M^R$, $M_0^R$, and~$M_-^R$.
Note that as $t$ is a global coordinate on the Lorentzian sections,
$\ttilde$~is a global coordinate on the Riemannian sections, and
$M^R$, $M_0^R$, and $M_-^R$ are
well defined.

$M_0^R$ and $M_-^R$ are the quotient spaces of $M^R$ with respect to
the Riemannian counterparts of the maps $J_0$ and
$J_-$~(\ref{jnoughtminus}). The coordinates $(\ttilde,x,y,z)$ are
global on~$M^R$, whereas on $M_0^R$ and $M_-^R$ they have the
identifications
\begin{mathletters}
\label{riem-flat-identifications}
\begin{eqnarray}
&&
(\ttilde,x,y,z) \sim (\ttilde,x,y,z+2a)
\ \ ,
\ \ \
\hbox{for $M_0^R$}
\ \ ,
\label{riemo-idents-Mnought}
\\
&&
(\ttilde,x,y,z) \sim (\ttilde,-x,-y,z+a)
\ \ ,
\ \ \
\hbox{for $M_-^R$}
\ \ .
\label{riemo-idents-Mminus}
\end{eqnarray}
\end{mathletters}
The metric reads explicitly
\begin{equation}
ds^2_R = d\ttilde^2 + dx^2 + dy^2 + dz^2
\ \ .
\label{riem-metric}
\end{equation}
The isometries of~$M^R$, $M_0^R$, and $M_-^R$ are
clear from the quotient construction.

We now wish to understand how the (local) Lorentzian Rindler
coordinates $(\eta,\xi,y,z)$, defined on the Rindler wedges of~$M$,
$M_0$, and~$M_-$, are continued into (local) Riemannian Rindler
coordinates on respectively~$M^R$, $M_0^R$, and~$M_-^R$. For $M$
and~$M_0$, the situation is familiar. Setting $t=-i\ttilde$ and $\eta
= -i\etatilde$, the transformation (\ref{rindler-coords}) becomes
\begin{mathletters}
\label{riem-rindler-coords}
\begin{eqnarray}
  \ttilde &=& \xi \sin(\etatilde)
  \ \ ,
  \\
  x &=& \xi \cos(\etatilde)
  \ \ ,
\end{eqnarray}
\end{mathletters}
and the metric (\ref{riem-metric}) reads
\begin{equation}
ds^2_R = \xi^2 d\etatilde^2 + d\xi^2 + dy^2 + dz^2
\ \ .
\label{riem-rindler-metric}
\end{equation}
On~$M^R$, one can therefore understand the set $(\etatilde,\xi,y,z)$
as (local) Riemannian Rindler coordinates, such that $\xi>0$ and
$\etatilde$ is periodically identified as $(\etatilde,\xi,y,z) \sim
(\etatilde+2\pi,\xi,y,z)$. The only part of $M^R$ not covered by
these coordinates is the flat $\BbbR^2$ of measure zero at $\xi=0$.
On~$M_0^R$, one has the additional identification
$(\etatilde,\xi,y,z) \sim (\etatilde,\xi,y,z+2a)$, which arises
from~(\ref{riemo-idents-Mnought}). On both $M^R$ and~$M_0^R$, the
globally-defined Killing vector $\partial_\etatilde = x
\partial_\ttilde - \ttilde \partial_x$ generates a $\Uone$ isometry
group of rotations about the origin in the $(\ttilde,x)$-planes. The
geometry is often described by saying that the Riemannian Rindler
time $\etatilde$ is periodic with period~$2\pi$, and the $\Uone$
isometry group is referred to as `translations in the Riemannian
Rindler time.'

On~$M_-^R$, we can again introduce by (\ref{riem-rindler-coords}) the
local Riemannian Rindler coordinates $(\etatilde,\xi,y,z)$ which,
with $\xi>0$, cover in local patches all of $M_-^R$ except the flat
open M\"obius strip of measure zero at $\xi=0$. The identifications
in these coordinates read
\begin{eqnarray}
(\etatilde,\xi,y,z)
&\sim&
(\etatilde+2\pi,\xi,y,z)
\nonumber
\\
&\sim&
(\pi - \etatilde,\xi,-y,z+a)
  \ \ ,
\end{eqnarray}
the latter one arising from~(\ref{riemo-idents-Mminus}). The
locally-defined Killing vector $\partial_\etatilde = x
\partial_\ttilde - \ttilde \partial_x$ can be extended into a smooth
line field ${\tilde V}^R$ (a vector up to a sign) on~$M_-^R$, but the
identification (\ref{riemo-idents-Mminus})
makes it impossible to promote ${\tilde
V}^R$ into a smooth vector field on $M_-^R$ by a consistent choice of
the sign. This means that $M_-^R$ does not admit a global $\Uone$
isometry that would correspond to `translations in the Riemannian
Rindler time'.

$M_-^R$ does, however, possess subsets that admit such $\Uone$
isometries. It is easy to verify that any point ${\bf x}\in M_-^R$
with $\xi>0$ has a neighborhood $U \simeq S^1 \times \BbbR^3$ with
the following properties:
1)~The restriction of ${\tilde V}^R$ to $U$ can be promoted into a
unique, complete vector field $V^R_U$ in $U$ by choosing the sign at
one point;
2)~The flow of $V^R_U$ forms a freely-acting $\Uone$ isometry group
of~$U$;
3) On~$U$, the Riemannian Rindler time $\etatilde$ can be defined as
an angular coordinate with period~$2\pi$, and the action of the
$\Uone$ isometry group on $U$ consists of `translations'
in~$\etatilde$. In this sense, one may regard $\etatilde$ on $M_-^R$
as a local angular coordinate with period~$2\pi$.

We can now turn to the Feynman propagators on our spacetimes.
Recall first that the Feynman propagator $G_M^F$
analytically continues into the Riemannian Feynman propagator
$G_{M^R}^F$, which depends on its two arguments only through the
Riemannian distance function
$\left[
(\ttilde-\ttilde')^2
+ (x-x')^2
+ (y-y')^2
+ (z-z')^2
\right]^{1/2}$,
and whose only singularity is at the coincidence limit.
$G_{M^R}^F$ is therefore invariant under the full isometry group
of~$M^R$. As the Riemannian Feynman propagators
on $M_0^R$ and $M_-^R$ are obtained
from $G_{M^R}^F$ by the method of
images, they are likewise invariant under the respective
full isometry groups of
$M_0^R$ and~$M_-^R$, and they are singular only at the coincidence
limit. In the massless case,
explicit expressions can be found by analytically continuing the
Lorentzian Feynman propagators given in
subsection~\ref{subsec:fields-on-flats}.

The properties of interest of the
Riemannian Feynman propagators can now be inferred from
the above discussion of the Riemannian Rindler coordinates.
It is immediate that
$G_{M^R}^F$ and $G_{M_0^R}^F$ are
invariant under the rotations generated by the
Killing vector~$\partial_\etatilde$, respectively on
$M^R$ and~$M_0^R$, and that they are periodic in $\etatilde$
in each argument with period~$2\pi$.
This periodicity of the propagator in
Riemannian time is characteristic of thermal
Green's functions. 
The local temperature seen by the observer
(\ref{observer-line}) is read off from the period
by relating $\eta$ to
the observer's proper time and the local redshift 
factor, with the result 
$T = {(2\pi\alpha)}^{-1}$
\cite{birrell-davies,takagi-magnum}.

$G_{M_-^R}^F$, on the other hand, displays no similar rotational
invariance. This is the Riemannian manifestation of the fact that
the restriction of $\minusvac$ to $R_-$ is not
invariant under the boost isometries of $R_-$ generated
by~$\partial_\eta$. $G_{M_-^R}^F$~is invariant under `local $2\pi$
translations' of each argument in~$\etatilde$, in the above-explained
sense in which $\etatilde$ provides on $M_-$ a local coordinate with
periodicity~$2\pi$. However, in the absence of a continuous
rotational invariance, it is difficult to
draw conclusions about the thermal character of 
$\minusvac$ merely by
inspection of the symmetries of~$G_{M_-^R}^F$.

One can, nevertheless, 
use the complex analytic properties of
$G_{M_-^R}^F$ to explicitly calculate the relation between the
quantum mechanical probabilities of the vacuum $\minusvac$ to emit
and absorb a Rindler particle with prescribed quantum numbers. We
shall briefly describe this calculation in
subsection~\ref{subsec:rpthree-path}, after having performed the
analogous calculation on the $\RPthree$ geon. For late and early
Rindler times, the emission and absorption probabilities of Rindler
particles with local frequency $E$ turn out to be related by the
factor $e^{-2\pi\alpha E}$. This is the characteristic thermal result
at the Rindler temperature $T = {(2\pi\alpha)}^{-1}$.

\section{The complexified Kruskal and
$\RPthree$ geon spacetimes}
\label{sec:kruskal-rpthree-defs}

This section is a mathematical interlude in which we describe the
Lorentzian and Riemannian sections of the complexified $\RPthree$
geon. The main point is to show how the quotient construction of the
Lorentzian $\RPthree$ geon from the Lorentzian Kruskal spacetime
\cite{topocen} can be analytically continued to the Riemannian
sections of the respective complexified manifolds. When formalized in
terms of (anti)holomorphic involutions on the complexified Kruskal
manifold \cite{gibb-holo,chamb-gibb}, this observation follows in a
straightforward way from the constructions of Ref.\
\cite{chamb-gibb}.

$M>0$ denotes throughout the Schwarzschild mass.

\subsection{Complexified Kruskal}
\label{subsec:kruskal-def}

Let $(Z^1,Z^2,Z^3,Z^4,Z^5,Z^6,Z^7)$ be global complex coordinates
on~$\BbbC^7$, and let $\BbbC^7$ be endowed with the flat metric
\begin{equation}
ds^2 =
{(dZ^1)}^2 +
{(dZ^2)}^2 +
{(dZ^3)}^2 +
{(dZ^4)}^2 +
{(dZ^5)}^2 +
{(dZ^6)}^2 -
{(dZ^7)}^2
\ \ .
\end{equation}
We define the complexified Kruskal spacetime ${\cal M^{\Bbb C}}$ as
the algebraic
variety in $\BbbC^7$ determined by the three polynomials
\cite{ferraris}
\begin{mathletters}
\begin{eqnarray}
(Z^{6})^{2} - (Z^{7})^{2} + \case{4}{3} (Z^{5})^{2}
&=&
16M^{2}
\ \ ,
\\
\left[(Z^{1})^{2} + (Z^{2})^{2} + (Z^{3})^{2}\right]
(Z^{5})^{4}
&=&
576M^{6}
\ \ ,
\\
\sqrt{3} Z^{4} Z^{5} + (Z^{5})^{2}
&=&
24M^{2}
\ \ .
\end{eqnarray}
\end{mathletters}
The Lorentzian and Riemannian sections of interest, denoted by
${\tilde{\cal M}}^L$ and~${\tilde{\cal M}}^R$, are the
subsets of ${\cal M^{\Bbb C}}$ stabilized by the respective
antiholomorphic involutions \cite{gibb-holo,chamb-gibb}
\begin{mathletters}
\begin{eqnarray}
&&{\cal J}_L:
(Z^1,Z^2,Z^3,Z^4,Z^5,Z^6,Z^7)
\mapsto
(\overline{Z^1},
\overline{Z^2},
\overline{Z^3},
\overline{Z^4},
\overline{Z^5},
\overline{Z^6},
\overline{Z^7})
\ \ ,
\\
&&{\cal J}_R:
(Z^1,Z^2,Z^3,Z^4,Z^5,Z^6,Z^7)
\mapsto
(\overline{Z^1},
\overline{Z^2},
\overline{Z^3},
\overline{Z^4},
\overline{Z^5},
\overline{Z^6},
-\overline{Z^7})
\ \ .
\end{eqnarray}
\end{mathletters}
${\tilde{\cal M}}^L$ and ${\tilde{\cal M}}^R$
are clearly real algebraic varieties.
On~${\tilde{\cal M}}^L$, $Z^i$ are real for all~$i$;
on~${\tilde{\cal M}}^R$, $Z^i$ are real for $1\le i \le 6$
while $Z^7$ is purely imaginary.

The Lorentzian section ${\tilde{\cal M}}^L$ consists of two connected
components, one with $Z^5>0$ and the other with $Z^5<0$. Each of these
components is isometric to the Kruskal spacetime, which we denote
by~${\cal M}^L$. An explicit embedding of ${\cal M}^L$ onto the
component of ${\tilde{\cal M}}^L$ with $Z^5>0$ reads, in terms of the
usual Kruskal coordinates $(T,X,\theta,\varphi)$,
\begin{mathletters}
\label{Kruskal-lor-embedding}
\begin{eqnarray}
Z^1
&=&
r \sin\theta \cos\varphi
\ \ ,
\label{Kruskal-lor-embedding1}
\\
Z^2
&=&
r \sin\theta \sin\varphi
\ \ ,
\\
Z^3
&=&
r \cos\theta
\ \ ,
\\
Z^4
&=&
4M \left(\frac{r}{2M}\right)^{1/2}
- 2M \left(\frac{2M}{r}\right)^{1/2}
\ \ ,
\\
Z^5
&=&
2M
\left(\frac{6M}{r}\right)^{1/2}
\ \ ,
\label{Kruskal-lor-embedding5}
\\
Z^6
&=&
4M
\left(\frac{2M}{r}\right)^{1/2}
\exp\left(-\frac{r}{4M}\right)
X
\ \ ,
\label{Kruskal-lor-embedding6}
\\
Z^7
&=&
4M
\left(\frac{2M}{r}\right)^{1/2}
\exp\left(-\frac{r}{4M}\right)
T
\ \ ,
\label{Kruskal-lor-embedding7}
\end{eqnarray}
\end{mathletters}
where $X^2 - T^2 > -1$, and $r$ is determined as a function of $T$
and $X$ from
\begin{equation}
\left(
{r \over 2M} -1
\right)
\exp \! \left(\frac{r}{2M}\right)
=
X^2 -T^2
\ \ .
\end{equation}
In the Kruskal coordinates, the metric on ${\cal M}^L$ reads
\begin{equation}
ds^2_{L} =
{32M^3 \over r} \exp\left(-\frac{r}{2M}\right)
\left( -dT^2 + dX^2 \right)
+ r^2 d\Omega^2
\ \ ,
\label{Kruskal-metric-lor}
\end{equation}
where $d\Omega^2 = d\theta^2 + \sin^2\theta d\varphi^2$ is the metric
on the unit two-sphere. In what follows, the singularities of the
spherical coordinates $(\theta,\varphi)$ on $S^2$ can be handled in
the standard way, and we shall 
not explicitly comment on these singularities. 

${\cal M}^L$~is 
both time and space orientable, and it admits
a global foliation with spacelike hypersurfaces whose topology is
$S^2 \times \BbbR \simeq S^3 \setminus${}$\{$two points at
infinity$\}$. ${\cal M}^L$~is manifestly spherically symmetric, with
an $\Othree$ isometry group that acts transitively on the two-spheres
in the metric~(\ref{Kruskal-metric-lor}). ${\cal M}^L$~has also the
Killing vector
\begin{equation}
V^L :=
\frac{1}{4M}
\left(
X \partial_T + T \partial_X
\right)
\ \ ,
\label{KillingV-lor}
\end{equation}
which is timelike for $|X| > |T|$ and spacelike for $|X| < |T|$. We
define the time orientation on ${\cal M}^L$ so that $V^L$ is
future-pointing for $X > |T|$ and past-pointing for $X < - |T|$.
A~conformal diagram of ${\cal M}^L$, with the two-spheres suppressed,
is shown in Figure~\ref{fig:Kruskal}.

In each of the four regions of ${\cal M}^L$ in which $|X| \neq |T|$,
one can introduce local Schwarzschild coordinates
$(t,r,\theta,\varphi)$ that are adapted to the isometry generated
by~$V^L$. 
In the exterior region $X > |T|$, this 
coordinate transformation reads
\begin{mathletters}
\label{Sch-lor-coords}
\begin{eqnarray}
T
&=&
\left(\frac{r}{2M}-1\right)^{1/2}
\exp \! \left(\frac{r}{4M}\right)
\sinh\left(\frac{t}{4M}\right)
\ \ ,
\\
X
&=&
\left(\frac{r}{2M}-1\right)^{1/2}
\exp \! \left(\frac{r}{4M}\right)
\cosh\left(\frac{t}{4M}\right)
\ \ ,
\end{eqnarray}
\end{mathletters}
where $r>2M$ and $-\infty<t<\infty$. The metric takes the 
familiar form
\begin{equation}
ds^2_{L} =
- \left(1 - \frac{r}{2M}\right) dt^2
+
{dr^2 \over
{\displaystyle{\left(1 - \frac{r}{2M}\right)}}}
+ r^2 d\Omega^2
\ \ ,
\label{Sch-metric-lor}
\end{equation}
and $V^L = \partial_t$.

The Riemannian section ${\tilde{\cal M}}^R$ consists of two connected
components, one with $Z^5>0$ and the other with $Z^5<0$. Each of
these components is isometric to the (usual) Riemannian Kruskal
spacetime, which we denote by~${\cal M}^R$. An explicit embedding of
${\cal M}^R$ onto the component of ${\tilde{\cal M}}^L$ with $Z^5>0$
is obtained, in terms of the usual Riemannian Kruskal coordinates
$({\tilde{T}},X,\theta,\varphi)$, by setting $T = -i {\tilde{T}}$ in
(\ref{Kruskal-lor-embedding})--(\ref{Kruskal-metric-lor}).
The ranges of ${\tilde{T}}$ and $X$ are unrestricted.

${\cal M}^R$ is 
orientable, and it admits an $\Othree$
isometry group that acts transitively on the two-spheres in the
Riemannian counterpart of the metric~(\ref{Kruskal-metric-lor}). It
also admits the Killing vector
\begin{equation}
V^R :=
\frac{1}{4M}
\left(
X \partial_{\tilde{T}} - {\tilde{T}} \partial_X
\right)
\ \ ,
\label{KillingV-riem}
\end{equation}
which is the Riemannian counterpart of $V^L$~(\ref{KillingV-lor}).
The Riemannian horizon is a two-sphere at ${\tilde{T}}=X=0$, where
$V^R$ vanishes.

With the exception of the Riemannian horizon, ${\cal M}^R$ can be
covered with the Riemannian Schwarzschild coordinates
$({\tilde{t}},r,\theta,\varphi)$, which are obtained from the
Lorentzian Schwarzschild coordinates in the region $X>|T|$ by setting
$t=-i {\tilde{t}}$ and taking ${\tilde{t}}$ periodic with
period~$8\pi M$. The well-known singularity of the Riemannian
Schwarzschild coordinates at the Riemannian horizon is that of
two-dimensional polar coordinates at the origin.

The intersection of ${\cal M}^L$ and ${\cal M}^R$ embeds into both
${\cal M}^L$ and ${\cal M}^R$ as a maximal three-dimensional wormhole
hypersurface of topology $S^2 \times \BbbR$. In the Lorentzian
(Riemannian) Kruskal coordinates, this hypersurface is given by $T=0$
(${\tilde{T}}=0$). 

\subsection{Complexified $\RPthree$ geon}
\label{subsec:rpthree-def}

Consider on ${\cal M^{\Bbb C}}$ the map \cite{chamb-gibb}
\begin{equation}
J:
(Z^1,Z^2,Z^3,Z^4,Z^5,Z^6,Z^7)
\mapsto
(-Z^1,-Z^2,-Z^3,Z^4,Z^5,-Z^6,Z^7)
\ \ .
\end{equation}
$J$~is clearly an involutive holomorphic
isometry,
and it acts freely on~${\cal M^{\Bbb C}}$.
We define the complexified $\RPthree$ geon spacetime
as the quotient space~${\cal M^{\Bbb C}}/J$.
In the notation of
Ref.\ \cite{chamb-gibb},
$J = {\cal R}_Z {\cal P}$.

As $J$ commutes with ${\cal J}_L$ and~${\cal J}_R$, the restrictions
of $J$ to ${\tilde{\cal M}}^L$ and ${\tilde{\cal M}}^R$ are
freely-acting involutive isometries. As $J$ leaves $Z^5$ invariant,
these isometries of ${\tilde{\cal M}}^L$ and ${\tilde{\cal M}}^R$
restrict further into isometries of each of the connected components.
$J$~thus restricts into freely and properly discontinuously acting
involutive isometries on ${\cal M}^L$ and~${\cal M}^R$. We denote
these isometries respectively by $J^L$ and~$J^R$. The Lorentzian
$\RPthree$ geon is now defined as the quotient space~${\cal
M}^L/J^L$, and the Riemannian $\RPthree$ geon is defined as the
quotient space~${\cal M}^R/J^R$. Their intersection is a
three-dimensional hypersurface of topology $\RPthree\setminus${}$\{$a
point at infinity$\}$, embedding as a maximal hypersurface into both
${\cal M}^L/J^L$ and~${\cal M}^R/J^R$.

For elucidating the geometries of 
${\cal M}^L/J^L$ and~${\cal M}^R/J^R$, it 
is useful to write the maps $J^L$ and $J^R$ in explicit coordinates.
In the Lorentzian (Riemannian) Kruskal coordinates on
${\cal M}^L$
(${\cal M}^R$, respectively),
we have
\begin{mathletters}
\begin{eqnarray}
&&J^L:
(T,X,\theta,\varphi)
\mapsto
(T,-X,\pi-\theta,\varphi+\pi)
\ \ ,
\label{JL-in-Kruskalcoords}
\\
&&J^R:
({\tilde{T}},X,\theta,\varphi)
\mapsto
({\tilde{T}},-X,\pi-\theta,\varphi+\pi)
\ \ .
\end{eqnarray}
\end{mathletters}
In the Riemannian Schwarzschild coordinates on~${\cal M}^R$,
$J^R$~reads
\begin{equation}
J^R:
({\tilde{t}},r,\theta,\varphi)
\mapsto
({\tilde{t}} + 4\pi M, r,\pi-\theta,\varphi+\pi)
\ \ .
\end{equation}
It is clear that $J^L$ preserves both time orientation and space
orientation
on~${\cal M}^L$, and $J^R$ preserves orientation on~${\cal M}^R$.
${\cal M}^L/J^L$~is therefore both time and space orientable, and
${\cal M}^R/J^R$ is orientable.

Consider first~${\cal M}^L/J^L$. As $J^L$ commutes with the $\Othree$
isometry of~${\cal M}^L$, ${\cal M}^L/J^L$ admits 
the induced
$\Othree$ isometry with two-dimensional spacelike orbits: ${\cal
M}^L/J^L$ is spherically symmetric. On the other hand, the Killing
vector $V^L$ of ${\cal M}^L$ changes sign under~$J^L$, and it
therefore induces only a line field $V^L/J^L$ but no globally-defined
vector field on~${\cal M}^L/J^L$. This means that ${\cal M}^L/J^L$
does not admit globally-defined isometries that would 
be locally generated by~$V^L/J^L$. 
Algebraically, this can be seen by noticing that
$J^L$ does not commute with the isometries of ${\cal M}^L$ 
generated by~$V^L$. 

A~conformal diagram of ${\cal M}^L/J^L$ is shown in
Figure~\ref{fig:RPthree}. Each point in the diagram represents an
$\Othree$ isometry orbit.
The region $X>0$ is isometric to that in
the Kruskal diagram of Figure~\ref{fig:Kruskal}, and the 
$\Othree$ isometry orbits are two-spheres.
At $X=0$, the $\Othree$ isometry orbits have topology~$\RPtwo$:
it is this set of exceptional orbits that 
cannot be consistently moved by the local isometries
generated by~$V^L/J^L$. 
${\cal M}^L/J^L$~is inextendible, and it admits
a global foliation with spacelike hypersurfaces whose topology is
$\RPthree\setminus${}$\{$a point at infinity$\}$.

${\cal M}^L/J^L$ is clearly an eternal black hole spacetime. It
possesses one asymptotically flat infinity, and an associated static
exterior region that is isometric to one Kruskal exterior region.
As mentioned above,
the exterior timelike Killing vector cannot be extended into a
global Killing vector on~${\cal M}^L/J^L$. 
Among the constant Schwarzschild time hypersurfaces in the
exterior region, there is only one that can be extended into a smooth
Cauchy hypersurface for~${\cal M}^L/J^L$: in
our (local) coordinates
$(T,X,\theta,\varphi)$, this distinguished exterior hypersurface is
at $T=0$.

The intersection of the past and future horizons is the two-surface on
which the Killing line field $V^L/J^L$ vanishes. This critical
surface has topology $\RPtwo$ and area~$8\pi M^2$. Away from the
critical surface, the future and past horizons have topology $S^2$ and
area~$16\pi M^2$, just as in Kruskal.

A~parallel discussion holds for~${\cal M}^R/J^R$. ${\cal
M}^R/J^R$~inherits from ${\cal M}^R$ an $\Othree$ isometry whose
generic orbits are two-spheres, but there is an exceptional
hypersurface of topology $\BbbR\times\RPtwo$ on which the orbits have
topology~$\RPtwo$. The `location' of this hypersurface prevents one
from consistently extending the local isometries generated by the
line field $V^R/J^R$ into globally-defined isometries. 

The line field $V^R/J^R$ can be promoted into a globally-defined
Killing vector field only in certain subsets of~${\cal M}^R/J^R$. In
particular, any point ${\bf x}\in {\cal M}^R/J^R$ with $r>2M$ has a
neighborhood $U \simeq S^1 \times \BbbR^3$ with the following
properties:
1)~The restriction of $V^R/J^R$ to $U$ can be promoted into a unique,
complete vector field $V^R_U$ in $U$ by choosing the sign at one
point;
2)~The flow of $V^R_U$ forms a freely-acting $\Uone$ isometry group
of~$U$;
3) On~$U$, the Riemannian Schwarzschild time ${\tilde{t}}$ can be
defined as an angular coordinate with period~$8\pi M$, and the action
of the $\Uone$ isometry group on $U$ consists of `translations'
in~${\tilde{t}}$. In this sense, one may regard ${\tilde{t}}$ on
${\cal M}^R/J^R$ as a local angular coordinate with period~$8\pi M$.

We define the Riemannian horizon as the set on which the Riemannian
Killing line field $V^R/J^R$ vanishes. This horizon is located at
$X=0=T$, and it is a surface with topology $\RPtwo$ and area $8\pi
M^2$ at $X=0=T$. The Riemannian horizon clearly lies in the
intersection of ${\cal M}^R/J^R$ and ${\cal M}^L/J^L$, and on ${\cal
M}^L/J^L$ it consists of the set where the Lorentzian
Killing line field $V^L/J^L$ vanishes. The Riemannian horizon thus
only sees the part of the Lorentzian horizon that is exceptional in
both topology and area. This will prove important for the
geon entropy in section~\ref{sec:rpthree-entropy}.

The above discussion is intended to emphasize the parallels between
the black hole spacetimes and the flat spacetimes of
section~\ref{sec:flats}. The Kruskal spacetime ${\cal M}^L$ is
analogous
to~$M_0$, and the $\RPthree$ geon ${\cal M}^L/J^L$ is analogous to
$M_-=M_0/{\tilde J}_-$. The isometries of ${\cal M}^L$ generated by
$V^L$ correspond to the
boost-isometries of $M_0$ generated by the Killing vector $t
\partial_x + x \partial_t$. The analogies of the conformal diagrams
in figures \ref{fig:Kruskal} and \ref{fig:RPthree} to those in
figures \ref{fig:Mnought} and \ref{fig:Mminus} are clear. The analogy
extends to the Riemannian sections of the flat spacetimes, discussed
in subsection~\ref{subsec:flat-imagtime}. The $\Uone$ isometry of
${\cal M}^R$ generated by $V^R$ corresponds to the $\Uone$ isometry
of $M^R_0$ generated by~$\partial_\etatilde$, and the $8\pi M$
periodicity of ${\tilde{t}}$ on ${\cal M}^R$ corresponds to the
$2\pi$ periodicity of $\etatilde$ on~$M^R_0$. The `local $8\pi M$
periodicity' of ${\tilde{t}}$ on ${\cal M}^R/J^R$ corresponds to the
`local $2\pi$ periodicity' of $\etatilde$ on~$M^R_-$, but in neither
case is this local periodicity associated with a globally-defined
$\Uone$ isometry.
Finally, the intersection of the future and past acceleration
horizons on
$M_-$ is exceptional both in topology and in 
what we might call the `formal area' (though the actual area is
infinite), 
and it is precisely this exceptional part of the Lorentzian horizon
that becomes the horizon of the Riemannian section.

\section{Scalar field theory on the $\RPthree$ geon}
\label{sec:field-on-rpthree}

In this section we analyze scalar field theory on the $\RPthree$ geon
spacetime. Subsection \ref{subsec:boul-vac} reviews the construction
of the Boulware vacuum $\boulvac$ in one exterior Schwarzschild
region. The Bogoliubov transformation between $\boulvac$ and the
Hartle-Hawking-like vacuum $\hhvacgeon$ is presented in
subsection~\ref{subsec:boul-hh-bogo}.
Subsection \ref{subsec:detector-in-rpthree} discusses briefly the
experiences of a particle detector in~$\hhvacgeon$, concentrating on
a detector that is in the exterior region of the geon and static with
respect to the timelike Killing vector of this region. Subsection
\ref{subsec:rpthree-path} derives the Hawking effect
from the complex analytic properties of the Feynman propagator
in~$\hhvacgeon$.

\subsection{Boulware vacuum}
\label{subsec:boul-vac}

We begin by reviewing the quantization of a real scalar field $\phi$
in one exterior Schwarzschild region.

As the Kruskal spacetime has vanishing Ricci scalar,
the curvature coupling term drops
out from the scalar field action~(\ref{scalar-action-gen}),
and the field equation reads
\begin{equation}
\left(
\nabla^a \nabla_a - \fieldmass^2
\right)
\phi
= 0
\ \ .
\label{field-eq-schw}
\end{equation}
In the exterior Schwarzschild metric in the Schwarzschild
coordinates~(\ref{Sch-metric-lor}), the field equation
(\ref{field-eq-schw}) can be separated with the ansatz
\begin{equation}
\phi =
{(4\pi\omega)}^{-1/2}
r^{-1} R_{\omega l}(r) e^{-i\omega t}
Y_{lm}(\theta,\varphi)
\ \ ,
\label{separ-ansatz}
\end{equation}
where $Y_{lm}$ are the spherical
harmonics.\footnote{We use the
Condon-Shortley phase convention (see for example Ref.\
\cite{arfken}), in which
$Y_{l(-m)}(\theta,\varphi) =
{(-1)}^m \overline{Y_{lm}(\theta,\varphi)}$
and
$Y_{lm}(\pi-\theta,\varphi+\pi) =
{(-1)}^l Y_{lm}(\theta,\varphi)$.}
The equation for the radial function $R_{\omega l}(r)$ is
\begin{equation}
0 =
\left[
\frac{d^2}{d {r^*}^2}
+ \omega^2
- \left( 1 - \frac{2M}{r} \right)
\left( \fieldmass^2 + \frac{l(l+1)}{r^2}
+ \frac{2M}{r^3}
\right)
\right]
R_{\omega l}
\ \ ,
\label{radial-eq}
\end{equation}
where $r^*$ is the tortoise coordinate,
\begin{equation}
r^* := r + 2M \ln \left( \frac{r}{2M} -1 \right)
\ \ .
\end{equation}
The (indefinite) inner product, evaluated on a hypersurface of
constant~$t$, reads
\begin{equation}
(\phi_1,\phi_2) :=
i \int_{S^2} \sin\theta \, d\theta d\varphi
\int_{-\infty}^{\infty}
r^2 dr^*
\,
\overline{\phi_1}
\,
\tensor{\partial}_t
\phi_2
\ \ .
\label{ext-inner-product}
\end{equation}

For presentational simplicity, we now set the field mass to zero,
$\fieldmass=0$. The case $\fieldmass>0$ will be discussed at the end
of subsection~\ref{subsec:boul-hh-bogo}\null.

The vacuum of positive frequency mode functions with respect to the
timelike Killing vector $\partial_t$ is called the Boulware vacuum
\cite{boulware-vac,fulling}.
A~complete orthonormal basis of mode
functions with this property is recovered from the separation
(\ref{separ-ansatz}) by taking $\omega>0$ and choosing,
for each~$l$,
for $R_{\omega l}$ a basis of solutions that are
$2\pi\delta$-orthonormal in
$\omega$ in the Schr\"odinger-type inner product
$\int_{-\infty}^\infty dr^*
\overline{R_1} R_2$.
We shall now make a convenient choice for
such an orthonormal basis.

For each $l$~and~$m$, it follows from standard one-dimensional
Schr\"odinger scattering theory \cite{messiahI,dunfordII} that the
spectrum for $\omega$ is continuous and consists of the entire
positive real line, and further that the spectrum has twofold
degeneracy. One way \cite{chris-full} to break this degeneracy and
obtain an orthonormal basis would be to choose for $R_{\omega l}$ the
conventional scattering-theory eigenfunctions $\roarrow{R}_{\omega
l}$ and $\loarrow{R}_{\omega l}$ whose asymptotic behavior as $r^*
\to \pm\infty$ is
\begin{mathletters}
\label{Rarrowdsef}
\begin{equation}
\roarrow{R}_{\omega l} \sim
\cases{
e^{i \omega r^*} + \roarrow{A}_{\omega l} e^{-i \omega r^*}
\ ,
&$r^* \to -\infty$
\ \ ,
\cr
B_{\omega l} e^{i \omega r^*}
\ ,
&$r^* \to \infty$
\ \ ,
\cr}
\end{equation}
\begin{equation}
\loarrow{R}_{\omega l} \sim
\cases{
B_{\omega l} e^{-i \omega r^*}
\ ,
&$r^* \to -\infty$
\ \ ,
\cr
e^{-i \omega r^*} + \loarrow{A}_{\omega l} e^{i \omega r^*}
\ ,
&$r^* \to \infty$
\ \ .
\cr}
\end{equation}
\end{mathletters}
The coefficients satisfy \cite{messiahI}
\begin{mathletters}
\label{scat-identities}
\begin{eqnarray}
&&
0<|B_{\omega l}| \le1
\ \ ,
\\
&&
\roarrow{A}_{\omega l} \overline{B_{\omega l}}
= -
\overline{\loarrow{A}_{\omega l}} B_{\omega l}
\ \ ,
\\
&&
{\left| \roarrow{A}_{\omega l} \right|}^2
=
{\left| \loarrow{A}_{\omega l} \right|}^2
=
1 - {\left| B_{\omega l} \right|}^2
\ \ .
\end{eqnarray}
\end{mathletters}
The modes involving $\roarrow{R}_{\omega l}$ are purely outgoing at
infinity, and those involving $\loarrow{R}_{\omega l}$ are purely
ingoing at the horizon. This basis would be especially useful if one
were to consider vacua that are not invariant under the time
inversion $t\to-t$ \cite{unruh-magnum}. For us, however, it will be
more transparent to use a basis in which complex conjugation is
simple. To this end, we introduce the solutions $R^\pm_{\omega l}$
for which
\begin{mathletters}
\label{Rplusminusdef}
\begin{equation}
\sqrt{2} \, R^+_{\omega l}
\sim
\sqrt{1 + \sqrt{1 - {\left| \roarrow{A}_{\omega l} \right|}^2 }}
\>
e^{i\omega r^*}
+
\frac{\roarrow{A}_{\omega l}
e^{-i\omega r^*}}
{
\sqrt{1 + \sqrt{1 - {\left| \roarrow{A}_{\omega l} \right|}^2 }}
}
\ \ \ \hbox{as $r^* \to -\infty$}
\ ,
\label{Rplusdef}
\end{equation}
and
\begin{equation}
R^-_{\omega l}
=
\overline{R^+_{\omega l}}
\ \ .
\label{Rplusminusconj}
\end{equation}
\end{mathletters}
Equations (\ref{Rarrowdsef}) and (\ref{Rplusminusdef}) define
$R^\pm_{\omega l}$ uniquely. Using the
identities~(\ref{scat-identities}), it is straightforward to verify
that the set $\left\{R^\pm_{\omega l}\right\}$ is
$2\pi\delta$-orthonormal in $\omega$ in the Schr\"odinger-type inner
product. Conversely, it can be verified that the Schr\"odinger-type
orthonormality and the complex conjugate relation
(\ref{Rplusminusconj}) determine these solutions uniquely up to an
overall phase.

We now take the complete orthonormal set of positive frequency
modes to be  $\left\{u_{\omega lm}^\epsilon\right\}$, where the
index $\epsilon$ takes the values~$\pm$ and
\begin{equation}
u_{\omega lm}^\epsilon
:=
e^{i(l+|m|)\pi/2}
{(4\pi\omega)}^{-1/2}
r^{-1} R^\epsilon_{\omega l} e^{-i\omega t}
Y_{lm}
\ \ .
\label{boulware-modes}
\end{equation}
The orthonormality relation reads
\begin{equation}
(u_{\omega lm}^\epsilon ,u_{\omega' l' m'}^{\epsilon'} )
=
\delta_{\epsilon \epsilon'}
\delta_{l l'}
\delta_{m m'}
\delta(\omega-\omega')
\ \ ,
\end{equation}
with the complex conjugates satisfying a similar relation with a
minus sign, and the mixed inner products vanishing.

The asymptotic behavior of $R^+_{\omega l}$ at infinity is
\begin{equation}
\frac{\sqrt{2} \>
\overline{B_{\omega l}}
\,
R^+_{\omega l}}
{
{\left| B_{\omega l} \right|}
}
\sim
\sqrt{1 +
\sqrt{1 - {\left| \roarrow{A}_{\omega l} \right|}^2 }}
\>
e^{i\omega r^*}
+
\frac{\overline{\loarrow{A}_{\omega l}}
\,
e^{-i\omega r^*}}
{
\sqrt{1 +
\sqrt{1 - {\left| \roarrow{A}_{\omega l} \right|}^2 }}
}
\ \ \ \hbox{as $r^* \to \infty$}
\ .
\label{Rplusaltdef}
\end{equation}
When $\left| \roarrow{A}_{\omega l} \right|\ll1$, equations
(\ref{Rplusdef}) and (\ref{Rplusaltdef}) show that $u_{\omega
lm}^+$ is mostly outgoing, with small ingoing scattering
corrections both at the horizon and at infinity. When $\left|
\roarrow{A}_{\omega l} \right|$ is not small, the relative
weights of the incoming and outgoing components in $u_{\omega
lm}^+$ become comparable, both at the horizon and at infinity.
Analogous statements hold for~$u_{\omega lm}^-$, with ingoing and
outgoing reversed.

We expand the quantized field as
\begin{equation}
\phi =
\sum_{\epsilon l m }
\int_0^\infty d\omega
\left[
b_{\omega lm}^\epsilon u_{\omega lm}^\epsilon
+
\left( b_{\omega lm}^\epsilon \right)^\dagger
\overline{u_{\omega lm}^\epsilon}
\right]
\ \ ,
\end{equation}
where $b_{\omega lm}^\epsilon$ and
$\left( b_{\omega lm}^\epsilon \right)^\dagger$
are the annihilation and creation operators associated with the
Boulware mode $u_{\omega lm}^\epsilon$. The Boulware vacuum
$\boulvac$ is defined by
\begin{equation}
b_{\omega lm}^\epsilon \boulvac = 0
\ \ .
\end{equation}

\subsection{Hartle-Hawking-like vacuum
and the Bogoliubov transformation}
\label{subsec:boul-hh-bogo}

In the Kruskal spacetime~${\cal M}^L$, the Hartle-Hawking vacuum
$\hhvackrus$ is the vacuum of mode functions that are positive
frequency with respect to the affine parameters of the
horizon-generating null geodesics \cite{hh-vacuum,israel-vacuum}.
As $\hhvackrus$ is invariant under the involution~$J^L$, it induces
a unique vacuum on the $\RPthree$ geon ${\cal M}^L/J^L$. We denote
this Hartle-Hawking-like vacuum on ${\cal M}^L/J^L$
by~$\hhvacgeon$. In terms of, say, the corresponding Feynman
propagators $G_{\rm K}^F$ on the Kruskal spacetime and $G_{\rm G}^F$
on the $\RPthree$ geon, this construction is given by the method of
images,
\begin{equation}
\label{mi}
G_{\rm G}^F(x,x') =
G_{\rm K}^F(x,x')
+
G_{\rm K}^F
\biglb(x,J^L(x') \bigrb)
\ \ .
\end{equation}
The arguments of the functions on the two sides of (\ref{mi})
represent points on respectively the $\RPthree$ geon and on the
Kruskal spacetime in the sense of local charts with identifications,
as with the flat spaces in subsection \ref{subsec:flat-spaces}
[cf.~(\ref{GMminus-image})].  A~complete set of the mode functions
whose vacuum is $\hhvacgeon$ can be recovered by forming from the
Kruskal Hartle-Hawking mode functions linear combinations that are
invariant under $J^L$ \cite{boersma}.

Several other characterizations of the state $\hhvacgeon$ can also be
given. In particular, $\hhvacgeon$
can be defined as the analytic continuation of the Green's
function on the Riemannian $\RPthree$ geon ${\cal M}^R/J^R$, and as
the vacuum of mode functions that are positive
frequency with respect to the affine parameters of the
horizon-generating null geodesics of the geon. The first of these
characterizations follows from the observation \cite{hh-vacuum}
that $G^F_{\rm K}$
analytically continues to the Riemannian Green's function on the
Riemannian Kruskal manifold ${\cal M}^R$ and
that the Green's functions $G^F_{\rm K^R}$
on ${\cal M}^R$ and $G^F_{\rm G^R}$ on ${\cal M}^R/J^R$ are
related by the Riemannian version
of~(\ref{mi}).
The resulting $G^F_{\rm G^R}$ is regular everywhere except at
the coincidence limit, and so analytically continues
to~$G^F_{\rm G}$. The second
characterization follows from the observation that the
modes constructed in \cite{boersma} (or, for example, the $W$-modes
below) have, when restricted to any generator of the geon horizon,
no negative frequency part with respect to the affine parameter 
along that generator.

We wish to find the Boulware-mode content of~$\hhvacgeon$. 
To this end, 
we recall that the Boulware-mode content of $\hhvackrus$
can be found by an analytic continuation argument
\cite{unruh-magnum,israel-vacuum} that is closely similar to the
analytic continuation argument used in finding the Rindler-mode
content of the Minkowski vacuum \cite{unruh-magnum}. 
In subsection \ref{subsec:flat-bogo} we adapted the Rindler
analytic continuation arguments from Minkowski space 
first to $M_0$ and then to~$M_-$. The analogy between
the quotient constructions $M_0 \to M_- = M_0/{\tilde J}_-$ and
${\cal M}^L \to {\cal M}^L/J^L$ makes it straightforward to adapt
our flat spacetime analytic continuation to the geon. 
One finds that a complete
orthonormal set of $W$-modes in the exterior region of ${\cal
M}^L/J^L$ is
$\left\{W_{\omega l m}^\epsilon\right\}$, where
\begin{equation}
W_{\omega l m}^\pm
:=
{1 \over \sqrt{2 \sinh(4\pi M\omega)}}
\left(
e^{2\pi M\omega} \,
u_{\omega lm}^\pm
+
e^{-2\pi M\omega} \,
\overline{u_{\omega l(-m)}^\mp }
\right)
\ \ .
\label{W-modes-geon}
\end{equation}

In analogy with~(\ref{minusvac-expanded}), we find
\begin{equation}
\hhvacgeon
=
\prod_{\omega l m}
\left(
{1 \over \cosh(r_\omega)}
\sum_{q=0}^\infty
\exp(-4\pi M \omega q)
|q\rangle_{\omega l m}^+ |q\rangle_{\omega l (-m)}^-
\right)
\ \ ,
\label{hhvacgeon-expanded}
\end{equation}
where
\begin{equation}
\tanh(r_\omega) = \exp(-4\pi M\omega)
\ \ ,
\end{equation}
and $|q\rangle_{\omega l m}^\epsilon$ denotes the normalized
state with $q$ excitations in the mode~$u_{\omega lm}^\epsilon$,
\begin{equation}
|q\rangle_{\omega l m}^\epsilon
:=
(q!)^{-1/2}
{\left[
\left( b_{\omega lm}^\epsilon \right)^\dagger
\right]}^q
\boulvac
\ \ .
\end{equation}
Thus, $\hhvacgeon$ contains Boulware modes in correlated pairs.
For any set of operators that only couple to one member of each
correlated Boulware pair, it is seen as in
subsection~\ref{subsec:flat-bogo} that the expectation values in
$\hhvacgeon$ are
thermal, and the temperature measured at the
infinity is the Hawking temperature, 
$T = {(8\pi M)}^{-1}$. However, for operators that do not have
this special form, the expectation values are not thermal.

The definition of $\hhvacgeon$ gives no reason to expect that the
restriction of $\hhvacgeon$ to the exterior region would be
invariant under Schwarzschild time translations. That the
restriction indeed is not invariant becomes explicit upon
decomposing the Boulware modes
$\left\{u_{\omega lm}^\epsilon\right\}$ into wave packets that are
localized in the Schwarzschild time. Using the functions
$\left\{h_{\rho\sigma}\right\}$~(\ref{hfd}),
we define such
packets by
\begin{equation}
u_{\rho\sigma lm}^\epsilon
:=
\int_0^\infty d\omega
\,
h_{\rho\sigma}(\omega)
\,
u_{\omega lm}^\epsilon
\ \ .
\label{geon-uloc-def}
\end{equation}
$u_{\rho\sigma lm}^\epsilon$ is localized in $\omega$ around
the value $\omega_\rho := (\rho+\casehalf)\epsilon_2$ with
width~$\epsilon_2$. When $r^*$ is so large that the asymptotic
form (\ref{Rplusaltdef}) holds, we see as in
subsection \ref{subsec:flat-bogo-packets} that
$u_{\rho\sigma lm}^\epsilon$ is approximately localized in $t$
around two peaks, situated at
$t = - 2\pi \epsilon_2^{-1} \sigma \pm r^*$,
with heights determined by the coefficients in~(\ref{Rplusdef}),
and each having width~$2\pi\epsilon_2^{-1}$.
In fact, the discussion is somewhat simplified 
by the massless nature of the current case and by the
asymptotic flatness of the geon.
Taking $\epsilon_2\ll1$ and proceeding as in
subsection~\ref{subsec:flat-bogo-packets}, we find
\begin{equation}
\hhvacgeon
\approx
\prod_{\rho\sigma l m}
\left(
{1 \over \cosh(r_{\omega_\rho})}
\sum_{q=0}^\infty
\exp(-4\pi M \omega_\rho q)
|q\rangle_{\rho\sigma l m}^+
|q\rangle_{\rho (-\sigma) l (-m)}^-
\right)
\ \ ,
\label{pac-hhvacgeon-expanded}
\end{equation}
where $|q\rangle_{\rho\sigma l m}^\epsilon$ denotes the normalized
state with $q$ excitations in the mode~$u_{\rho\sigma
lm}^\epsilon$,
\begin{equation}
|q\rangle_{\rho\sigma l m}^\epsilon
:=
(q!)^{-1/2}
{\left[
\left( b_{\rho\sigma lm}^\epsilon \right)^\dagger
\right]}^q
\boulvac
\ \ .
\end{equation}
The noninvariance of $\hhvacgeon$ under Schwarzschild time
translations is apparent from the noninvariance of
(\ref{pac-hhvacgeon-expanded}) under (integer) translations
in~$\sigma$.

Consider now an observer in the exterior region at a constant value
of $r$ and the angular coordinates. At early (late) Schwarzschild
times, the mode localization properties discussed above imply
that the observer only couples to modes with large positive
(negative) values of~$\sigma$, and thus sees $\hhvacgeon$ as a
thermal state. In particular, the observer cannot distinguish
$\hhvacgeon$ from $\hhvackrus$ in these limits. For $r\gg M$, the
temperature is the Hawking temperature $T = {(8\pi M)}^{-1}$.

Just as in the flat space case, the correlations exhibited in
(\ref{pac-hhvacgeon-expanded}) should not be surprising. In the
vacuum $\hhvackrus$ in the Kruskal spacetime, invariance under
Killing time translations implies that the partner of a
right-hand-side Boulware mode localized at asymptotically early
(late) Schwarzschild times is a left-hand-side Boulware mode
localized at asymptotically late (early) Schwarzschild times. The
properties of the involution $J^L$ on the Kruskal spacetime lead one
to expect in $\hhvacgeon$ a correlation between Boulware modes at
early and late times, and a correlation between Boulware modes with
opposite signs of~$m$: this is indeed borne out
by~(\ref{pac-hhvacgeon-expanded}).

For the number operator expectation value of the
mode $u_{\rho\sigma l m}$ in~$\hhvacgeon$, one finds precisely the
same result as in~$\hhvacgeon$,
\begin{equation}
N_{\rho\sigma lm}
\approx
{\left( e^{8\pi M\omega_\rho} - 1 \right)}^{-1}
\ \ .
\label{gapp-loc-number-planckian}
\end{equation}
This is the Planckian distribution in the temperature 
$T = {(8\pi M)}^{-1}$. 
In particular, the number operator expectation value
contains no information about the noninvariance of $\hhvacgeon$
under the Schwarzschild time translations.

To end this subsection, we note that the
above discussion can be easily generalized to a scalar
field with a positive mass~$\fieldmass$. For each $l$~and~$m$, the
spectrum for $\omega$ is again continuous and consists of the entire
positive real line, but the spectrum is now degenerate only for
$\omega>\fieldmass$. In the nondegenerate part,
$0<\omega<\fieldmass$, the eigenfunctions vanish exponentially at
$r^*\to\infty$, while at $r^*\to-\infty$ they are asymptotically
proportional to $\cos(\omega r^* +\delta_{\omega l})$, where
$\delta_{\omega l}$ is a real phase. The nondegenerate part of the
spectrum thus corresponds classically to particles that never reach
infinity, and the Bogoliubov transformation for these modes is
qualitatively similar to that of the $n=0$ modes in
subsection~\ref{subsec:flat-bogo}. In the degenerate part of the
spectrum, $\omega>\fieldmass$, the asymptotic solutions to the
radial equation (\ref{radial-eq}) at $r^*\to\infty$ are now linear
combinations of ${(r^*/M)}^{\pm i \fieldmass^2 M/p} \exp(\pm ip
r^*)$, where $p := \sqrt{\omega^2 - \fieldmass^2}$, and the
relations (\ref{Rarrowdsef}) and (\ref{scat-identities})
need to be modified accordingly, but equations
(\ref{Rplusminusdef}) and (\ref{boulware-modes}) do then again
define an orthonormal set of modes, and the rest of the discussion
proceeds as in the massless case.
Thus, also in the massive case,
expectation values of operators that couple only to one member of
each correlated Boulware mode pair are thermal in the Hawking
temperature $T = {(8\pi M)}^{-1}$. 
Again, arguments similar to those of 
subsection \ref{subsec:flat-bogo-packets} 
show that this is the
case for any operators that only couple 
to the infinity-reaching modes at large distances and at
asymptotically early or late Schwarzschild times.

\subsection{Particle detector in the Hartle-Hawking-like vacuum}
\label{subsec:detector-in-rpthree}

We shall now briefly consider the response of a particle detector
on the $\RPthree$ geon in the vacuum~$\hhvacgeon$.

We describe the internal degrees of freedom of the detector by an
idealized monopole interaction as in
subsection~\ref{subsec:flat-detector}. In first order perturbation
theory, the detector transition probability is given by formulas
(\ref{exci-prob}) and~(\ref{resp-function}), where $x(\tau)$ is the
detector trajectory parametrized by the proper time, and $G^+(x,x')$
stands for the Wightman function $G_{\rm G}^+(x,x'):= \hhvacgeonbra
\phi(x) \phi(x') \hhvacgeon$. In analogy with~(\ref{mi}), we have
\begin{equation}
G_{\rm G}^+(x,x') =
G_{\rm K}^+(x,x')
+
G_{\rm K}^+
\biglb(x,J^L(x') \bigrb)
\ \ ,
\label{GplusG-image}
\end{equation}
where $G_{\rm K}^+(x,x'):= \hhvackrusbra \phi(x) \phi(x') \hhvackrus$
is the Kruskal Wightman function.

Of particular interest is a detector that is in the exterior region
and static with respect to the Schwarzschild time translation Killing
vector of this region. The contribution to the response function
(\ref{resp-function}) from the first term on the right-hand-side of
(\ref{GplusG-image}) is then exactly as in Kruskal, and this
contribution indicates a thermal response at the Hawking temperature
$T = {(8\pi M)}^{-1}$ \cite{unruh-magnum}. The new effects on the
$\RPthree$ geon are due to the additional contribution from
\begin{equation}
\Delta G^+_{\rm G}(\tau,\tau')
:=
G_{\rm K}^+
\Biglb( x(\tau), J^L \biglb( x(\tau') \bigrb)  \Bigrb)
\ \ .
\label{Delta-G-geon}
\end{equation}
Unfortunately, the existing literature on the Kruskal Wightman
functions seems to contain little information 
about~$\Delta G^+_{\rm G}$.
The points $x(\tau)$ and $J^L \biglb( x(\tau') \bigrb)$ in
(\ref{Delta-G-geon}) are in the opposite exterior Kruskal regions,
and field theory on the Kruskal spacetime gives little incentive to
study
the Wightman functions in this domain. We therefore only offer some
conjectural remarks.

As translations in the exterior Killing time cannot be extended into
globally-defined isometries of the $\RPthree$ geon, there is no
apparent symmetry that would make the detector excitation rate
independent of the proper time along the trajectory. However, from
the locations of the points $x(\tau)$ and $J^L \biglb( x(\tau')
\bigrb)$ in the Kruskal spacetime, it seems likely that $\Delta
G^+_{\rm G}(\tau,\tau')$ tends to zero when 
$|\tau+\tau'|$ tends to infinity, as was 
the case with the analogous quantity
(\ref{Delta-Gplus}) in the Rindler analysis on~$M_-$. If true, this
means that the responses in $\hhvacgeon$ and $\hhvackrus$ are
identical for a detector that only operates in the asymptotic past or
asymptotic future. Further, it seems likely that $G_{\rm K}^+(x,x')$
tends to zero whenever the points $x$ and $x'$ tend to large values
of the curvature radius in the opposite Kruskal exterior regions, as
a power law for a massless field and exponentially for a massive
field.\footnote{We thank Bob Wald for this remark.} If true, this
implies that the responses in $\hhvacgeon$ and $\hhvackrus$ become
asymptotically identical for a detector far from the hole, even if
the detector operates at proper times that are not in the asymptotic
past or future.

Finally, we note that the contribution to the renormalized
expectation value
$\hhvacgeonbra T_{\mu\nu}(x)\hhvacgeon$
from the second term in (\ref{GplusG-image})
is manifestly finite. If $G_{\rm K}^+(x,x')$ satisfies the falloff
properties mentioned above, and if its derivatives fall off
similarly, it follows that
$\hhvacgeonbra T_{\mu\nu}(x)\hhvacgeon$
approaches
$\hhvackrusbra T_{\mu\nu}(x)\hhvackrus$
when the curvature radius of the point $x$ is asymptotically large,
or when the point $x$ is taken to asymptotically distant future or
past along a path of fixed curvature radius in the exterior region.
If true, this means that the asymptotic agreement of the detector
responses in $\hhvacgeon$ and $\hhvackrus$ is accompanied by the
asymptotic agreement of the stress-energy tensors.

\subsection{Derivation of the Hawking effect from
the analytic properties of the Feynman propagator}
\label{subsec:rpthree-path}

In this subsection we derive the Hawking effect on the $\RPthree$
geon from the analytic properties of the Feynman propagator in the
vacuum~$\hhvacgeon$. The idea is to consider the probabilities of the
geon to emit and absorb particles with a given frequency, at late
exterior times, and to reduce these probabilities to those of the
Kruskal hole. This subsection is meant to be read in
close conjunction with the Kruskal analysis of Ref.\
\cite{hh-vacuum}.

Following section IV of Ref.\ \cite{hh-vacuum}, we envisage a family
of particle detectors located on a timelike hypersurface $O$ of
constant curvature radius in the exterior region of the $\RPthree$
geon (see Figure~\ref{fig:RPthree-em-abs}). The detectors are assumed
to measure particles that are purely positive frequency with respect
to the exterior Killing vector~$\partial_t$. The amplitude that a
particle is detected in a mode~$f_i(x')$, having started in a mode
$h_j(x)$ on some hypersurface $\tilde{O}$
that bounds a region interior
to~$O$, is given by equation (4.1) of Ref.\ \cite{hh-vacuum},
\begin{equation}
-
\int_{O} d\sigma^\mu(x')
\int_{\tilde{O}} d\sigma^\nu(x)
\,
\overline{f_i}(x')
\,
\tensor{\partial_\mu}
\,
G_{\rm G}^F (x',x)
\,
\tensor{\partial_\nu}
\,
h_j (x)
\ \ .
\label{hh-amplitude}
\end{equation}
If the mode $f_i(x')$ is peaked at an asymptotically late
Schwarzschild time, we argue as in Ref.\ \cite{hh-vacuum} that
$\tilde{O}$ can be replaced by
a spacelike hypersurface ${\tilde C}_+$
of constant two-sphere curvature radius in the black hole
interior.\footnote{Ref.\ \cite{hh-vacuum} invoked the Killing time
translation invariance of $G_{\rm K}^F (x',x)$ to argue that the
contribution from a timelike hypersurface connecting a point on $O$
to a point on the Kruskal counterpart of ${\tilde C}_+$ can be
neglected, by taking this timelike hypersurface to be in the distant
future. For us, this argument shows that one can neglect the
contribution from the first term on the right-hand-side
of~(\ref{mi}). To argue that the contribution from the
second term on the right-hand-side of (\ref{mi}) can be
neglected, again by taking the timelike hypersurface to be in the
distant future, it would be sufficient to show that $G_{\rm K}^F$
satisfies a slightly stronger falloff than that assumed for $G_{\rm
K}^+$ in subsection \ref{subsec:detector-in-rpthree}, which seems
likely to be the case.}
To find the total probability that a particle is detected, one thus
needs to compute the modulus squared of the amplitude
(\ref{hh-amplitude}) and sum over a complete set of states
$\left\{h_j\right\}$ on~${\tilde C}_+$.

Recall that in the future interior region on the Kruskal spacetime,
one can introduce the interior Schwarzschild coordinates
$(t,r,\theta,\varphi)$, in which the metric is given by
(\ref{Sch-metric-lor}) with $0<r<2M$, the coordinate $r$ decreases
toward the future, and the map $J^L$ (\ref{JL-in-Kruskalcoords})
reads
\begin{equation}
J^L:
(t,r,\theta,\varphi)
\mapsto
(-t, r,\pi-\theta,\varphi+\pi)
\ \ .
\label{JL-in-int-Sch}
\end{equation}
The interior Schwarzschild coordinates therefore provide in the
future interior region of the geon a set of local coordinates with
the identification $(t,r,\theta,\varphi) \sim
(-t, r,\pi-\theta,\varphi+\pi)$.
Working in this chart, 
we obtain a complete set of states $\left\{h_j\right\}$
on ${\tilde C}_+$ by separation of variables: the states are
proportional to
\begin{equation}
r^{-1} {\tilde R}_{\omega l}
\left[
e^{i\omega t} + {(-1)}^l e^{-i\omega t}
\right]
Y_{lm}
\ \ ,
\label{inter-separ-ansatz}
\end{equation}
where $\omega>0$
and ${\tilde R}_{\omega l}$ satisfies the
counterpart of equation (\ref{radial-eq}) for the interior region.

Consider now the integration over $x$
in~(\ref{hh-amplitude}),
with $\tilde{O}$ replaced by~${\tilde C}_+$.
In our coordinates, $r$~is constant on~${\tilde C}_+$,
and we cover ${\tilde C}_+$ precisely once,
up to a set of measure zero, by taking $t>0$
and letting the angles range over the full two-sphere. We write
$G_{\rm G}^F$ as in (\ref{mi}) and perform in the
second term the change of variables $t\to-t$. The amplitude
(\ref{hh-amplitude}) takes then the form
\begin{equation}
-
\int_{O} d\sigma^\mu(x')
\int_{C_+} d\sigma^\nu(x)
\,
\overline{f_i}(x')
\,
\tensor{\partial_\mu}
\,
G_{\rm K}^F (x',x)
\,
\tensor{\partial_\nu}
\,
h_j (x)
\ \ ,
\label{hh-amplitude-K}
\end{equation}
where the integration over $x$ is now over the
hypersurface $C_+$ of
constant $r$ in the future interior region of the {\em Kruskal\/}
spacetime (see Figure 3 of Ref.\ \cite{hh-vacuum}),
and the function $h_j$ has been extended into all of this
region by the formula~(\ref{inter-separ-ansatz}).

Let now the exterior mode function $f_i$ be of the form
(\ref{separ-ansatz}) with frequency $\omega'>0$. The invariance of
$G_{\rm K}^F$ under the Killing time translations on the Kruskal
spacetime implies that the integrals over $t$ and $t'$ in
(\ref{hh-amplitude-K}) yield a linear combination of two terms,
proportional respectively to $\delta(\omega-\omega')$ and
$\delta(\omega+\omega')$. The latter term is however vanishing,
because in it the argument of the delta-function is always positive.

These manipulations have reduced our amplitude to that analyzed in
Ref.\ \cite{hh-vacuum}. A~similar reduction can be performed for the
amplitude for a particle to propagate {\em from\/} the hypersurface
$O$ {\em to\/} the hypersurface ${\tilde C}_-$ of constant curvature
radius in the past interior region (see
Figure~\ref{fig:RPthree-em-abs}). The relation derived in Ref.\
\cite{hh-vacuum} for the Kruskal amplitudes, from the analytic
properties of~$G_{\rm K}^F$, 
holds therefore also for our amplitudes. 
We infer that the probability for the $\RPthree$ geon to
emit a late time particle with
frequency $\omega$ is $e^{-8\pi M\omega}$
times the probability for the geon to absorb a particle in the same
mode. This is the thermal result, at the Hawking temperature 
$T = {(8\pi M)}^{-1}$.

It should be emphasized that this derivation of the thermal spectrum
for $\hhvacgeon$ explicitly assumes that the emitted and absorbed
particles are in the distant future. By the global time reversal
invariance of the geon, the thermal result also holds for particles
that are emitted and absorbed in the distant past. It seems more
difficult to assess whether the result could be extended to particles
at finite values of the exterior Schwarzschild time, however. One
concern with such particles is whether one can justify the arguments
for choosing the interior hypersurfaces to be ${\tilde C}_+$ and
${\tilde C}_-$. Another concern is whether one would need to replace
the energy eigenstates (\ref{separ-ansatz}) by exterior modes that
are explicitly localized in the exterior Schwarzschild time. Note
that for $\hhvackrus$ neither of these concerns arise, as there the
Killing time translation symmetry implies that the thermal result
holds for particles emitted and absorbed at arbitrary values of the
exterior Schwarzschild time.

Finally, we remark that a similar emission-absorption analysis can be
performed for Rindler particles in the vacuum $\minusvac$ on~$M_-$.
In this case, the necessary assumptions about the falloff of the
Feynman propagator can be explicitly verified. For late and early
Rindler times, one finds that the vacuum emission and absorption
probabilities of Rindler particles with local frequency $E$ are
related by the thermal factor $e^{-2\pi\alpha E}$. This is the
thermal result at the Rindler temperature $T = {(2\pi\alpha)}^{-1}$.

\section{Entropy of the $\RPthree$ geon?}
\label{sec:rpthree-entropy}

We have seen that the Hartle-Hawking-like vacuum $\hhvacgeon$ on the
$\RPthree$ geon has certain characteristics of a thermal bath at the
Hawking temperature $T = {(8\pi M)}^{-1}$. We shall now discuss
whether it is possible to associate with the geon also a
gravitational entropy.

Consider first an observer $Q$ in the exterior Schwarzschild region.
The future of $Q$ may or may not be
isometric to a region of the Kruskal
spacetime, but the possible differences are hidden behind the black
hole horizon. For any classical means that $Q$ may employ to probe
the spacetime, such as letting matter fall into the black hole, the
response of the spacetime is to $Q$  indistinguishable from that of
the Kruskal spacetime, provided $Q$ remains outside the black hole
horizon also in the deformed spacetime. In particular, if $Q$ is in
the asymptotically flat region, and if $Q$ deforms the hole only in
the mass (but not in the angular momentum or charge), 
the first law of black hole mechanics takes for $Q$ 
the standard form \cite{BarCarHaw}
\begin{equation}
dM =
\frac{1}{32 \pi M} \, d A
\ \ ,
\label{mech-first-law}
\end{equation}
where
$A := 16 \pi M^2$ is the horizon area of a Kruskal hole with
mass~$M$. As discussed in subsection~\ref{subsec:rpthree-def}, $A$~is
equal to the geon horizon area away from the critical surface at the
intersection of the past and future horizons.

Suppose now that the quantum state is $\hhvacgeon$, and that $Q$ is
at late exterior Schwarzschild time. We have argued that $Q$ then
sees the black hole as being in equilibrium with a thermal bath at
the Hawking temperature $T = {(8\pi M)}^{-1}$. By the usual arguments
\cite{hawkingCMP,bekenstein1,bekenstein2}, this leads $Q$ to
reinterpret equation (\ref{mech-first-law}) as the first law of
thermodynamics,
\begin{equation}
dE =
T d S
\ \ ,
\label{ther-first-law}
\end{equation}
where $S = \case{1}{4}A$ is the entropy of the geon. This
entropy is exactly the same as in the Kruskal
spacetime with the same mass. 

Consider then the path-integral approach. Following Refs.\
\cite{GH1,hawkingCC}, we assume that the thermodynamics seen by an
observer at infinity is described by the partition function
\begin{equation}
Z(\beta) = \int {\cal D} g_{\mu\nu} \exp(-I)
\ \ ,
\label{partit-def}
\end{equation}
where $\beta$ is the inverse temperature at infinity, $I$~is the
action of the Riemannian metric~$g_{\mu\nu}$, and the boundary
conditions for the path integral are to encode the topology of the
manifold, the asymptotic flatness, the lack of angular momentum, and
the value of~$\beta$. We further assume that the partition function
can be estimated by the saddle point contribution,
\begin{equation}
Z(\beta) \approx \exp(-I^c)
\ \ ,
\label{partit-estimate}
\end{equation}
where $I^c$ is the action of the classical solution satisfying the
boundary conditions of the integral in~(\ref{partit-def}). Discussing
the validity of these assumptions at any general level falls beyond
the scope of this paper (for some perspectives, see for example
Refs.\ \cite{hawkingCC,york1,WYprl,hallhartle-con,JJHlouko3}), but
what we do wish to do is to contrast the consequences of these
assumptions for the Kruskal black hole and the $\RPthree$ geon.

When the Lorentzian thermodynamic object is the Kruskal hole, the
boundary conditions for the integral in (\ref{partit-def}) were
chosen in Ref.\ \cite{GH1} so that that the saddle point solution is
the Riemannian section of the Kruskal manifold, and $\beta$ was
identified with the period of the Riemannian Schwarzschild time. This
leads to $\beta = 8\pi M$, which reproduces the Hawking temperature.
To arrive at a finite action, one introduces a boundary with topology
$S^1 \times S^2$, subtracts from the action at this boundary a
boundary term that makes the action of flat space vanish, and then
lets the boundary go to infinity. The result is
\begin{equation}
I_K^c = 4\pi M^2 =
\frac{\beta^2}{16\pi}
\ \ .
\label{krus-riem-action}
\end{equation}
Using (\ref{partit-estimate}) and (\ref{krus-riem-action}) in the
formula for the entropy in the canonical ensemble,
\begin{equation}
S=\bigl[1 - \beta(\partial/\partial\beta)\bigr]
\ln Z
\ \ ,
\label{can-entropy-formula}
\end{equation}
one finds the Bekenstein-Hawking result, $S \approx \case{1}{4}A$.

When the Lorentzian thermodynamic object is the $\RPthree$ geon, it
seems reasonable to choose the boundary conditions for the integral
in (\ref{partit-def}) so that that the saddle point solution is the
Riemannian $\RPthree$ geon. The thermodynamics on the geon, discussed
in section~\ref{sec:field-on-rpthree}, then suggests introducing the
mass-temperature relation $\beta = 8\pi M$, which geometrically means
identifying $\beta$ with the `local period' of the Riemannian
Schwarzschild time on the Riemannian geon. To recover an action, we
note that the $S^1 \times S^2$ boundary prescription of Ref.\
\cite{GH1} for the Riemannian Kruskal manifold is invariant under the
map $J^R$ of subsection~\ref{subsec:rpthree-def}, and this
prescription therefore induces on the Riemannian geon a boundary
prescription that yields a finite action when the boundary is taken
to infinity. Proceeding via these steps, we find for the action of
the geon the result
\begin{equation}
I_G^c = 2\pi M^2 =
\frac{\beta^2}{32\pi}
\ \ ,
\label{geon-riem-action}
\end{equation}
which is half the Kruskal action~(\ref{krus-riem-action}). For the
entropy of the geon we then obtain, using~(\ref{partit-estimate}),
(\ref{can-entropy-formula}), and~(\ref{geon-riem-action}),
$S \approx\case{1}{8}A$. This is only half of the
Bekenstein-Hawking result for the Kruskal hole.

{}From the mathematical point of view, the relative factor
$\casehalf$ in the geon entropies obtained by the late-time
thermodynamic
arguments and the path-integral method should not be
surprising. The horizon area relevant for the
thermodynamic
arguments is that at late times along the future horizon, and this
area is exactly the same as in Kruskal. 
The horizon area underlying the path-integral entropy, 
on the other hand, is that of the Riemannian horizon, 
and we saw in
subsection \ref{subsec:rpthree-def} 
that this is only half of the
area of the Riemannian Kruskal horizon.

Physically, however, the disagreement between the two entropies
calls for an explanation. The physical framework of the first
derivation, via the observer $Q$ and the classical first
law~(\ref{mech-first-law}), is relatively clear, and it is
difficult to escape the conclusion that the Bekenstein-Hawking
entropy must be the correct one from the thermodynamic viewpoint of
the observer~$Q$. The framework of the path-integral derivation,
however, invites more scrutiny.

A~first possibility is that the path-integral framework is simply
inapplicable to the geon, for example due to the lack of sufficient
symmetry in any of several aspects of our discussion.
Recall, for instance, that despite the fact that the
exterior region of the spacetime is static,
the restriction of $\hhvacgeon$ to this region is not.
It seems likely that any state that is static in the exterior region
of the geon must become singular somewhere on the horizon.
Certainly, this is the case if one attempts the following
construction: Suppose that we identify the exterior of the geon
with an exterior region in Kruskal and consider the restriction of a
Feynman Green's function
$G_{\rm K}(x,x')$ of some static state to this region.  Such a
function cannot be smoothly extended to a regular Green's function
on the geon because it will not have the right singularities on the
bifurcation surface. Approaching the bifurcation surface with two
exterior  points $x$ and $x'$ on opposite sides of the two-sphere,
the Green's function $G_{\rm K} (x,x')$ will have a smooth limit on
the bifurcation surface.  In the geon spacetime, however,
both $x$ and $x'$ will approach the same point, and the Green's
function should diverge.

The saddle point solution that was used to arrive at
(\ref{geon-riem-action}) above is another object with rather less
symmetry than one might like. This saddle point solution is the
Riemannian $\RPthree$ geon ${\cal M}^R/J^R$, and it differs from the
Riemannian Kruskal manifold in both its metric and topological
properties. For example, the asymptotic region of ${\cal M}^R/J^R$
does not have a global Killing field, and the homotopy group of any
neighborhood of infinity in ${\cal M}^R/J^R$ is $\BbbZ_2$ as opposed
to the trivial group.  It may well be that such an asymptotic
structure does not satisfy the boundary conditions that should be
imposed in the integral~(\ref{partit-def}).\footnote{We thank Rafael
Sorkin for stressing the possible importance of the asymptotic
topology even in cases (unlike ours) in which the asymptotic
metric would admit a global Killing symmetry.}
We note that this point is connected to the one above as it shows
that no analytic state on the geon can be static in the exterior
region.

In the context of this discussion it is interesting to recall that
while the Riemannian $\RPthree$ geon has the asymptotic structure
just described, the single
asymptotic region of the Lorentzian $\RPthree$
geon is just the familiar one, that is, the asymptotic region of one
Kruskal exterior.  Thus, on the geon, the structure of the Riemannian
infinity is influenced not only by the structure of the single
Lorentzian infinity, but also by 
what lies behind the Lorentzian horizons. 

Another possibility is that the path-integral framework is applicable
to the geon, but that the proper procedure is more subtle than the one
outlined above.  However, it is difficult to see what reasonable
modification of the above steps would lead to a result consistent with
the first law.

A final possibility is that the path-integral
framework is applicable to
the geon, and our way of applying it is correct, but the resulting
entropy is physically distinct from the subjective thermodynamic
entropy associated with the observer~$Q$. If this is the case, the
physical interpretation of the path-integral entropy might be found
in the quantum statistics in the whole exterior region, rather than
just the thermodynamics of late times in the exterior region. In an
operator formalism, one might anticipate such an entropy to arise
from tracing over the degrees of freedom that are in some sense
unobservable.

{}From the operator point of view, the factor $\casehalf$ in the geon
entropy might even appear reasonable. In the Hartle-Hawking vacuum on
the Kruskal manifold,
the thermal expectation values for operators in one
exterior region arise from tracing over all the Boulware modes in the
second, unobserved exterior region. On the geon, on the other hand,
thermal expectation values in the Hartle-Hawking-like vacuum
$\hhvacgeon$ arise (for all exterior Schwarzschild times) only for
operators that do not couple to, and hence lead to a trace over, half
of the Boulware modes in the single exterior region. On the geon, the
thermal expectation values thus involve a trace over half as many
Boulware modes as on the Kruskal spacetime. If the statistical
entropy were somehow to count modes that are traced over 
in these expectation
values, the geon entropy indeed {\em should\/} be half of the Kruskal
entropy.\footnote{We thank Karel Kucha\v{r} for suggesting (but not
advocating) this argument.} An uncomfortable aspect of this argument
is, however, that the entropy $\case{1}{8}A$ would then reflect not
just the geometry of the geon and the properties of the
state~$\hhvacgeon$, but also the choice of a particular class of
operators in the exterior region, and it seems difficult to motivate
this choice on geometrical grounds only.

To end this section, we note that an analogous discussion can be
carried out for the entropy associated with the Rindler observer and
the acceleration horizon in the flat spacetimes $M_0$ and~$M_-$. The
horizon areas are formally infinite, owing to the infinite range of
the coordinate~$y$, but this appears to be a minor technicality: the
thermodynamic discussion of section \ref{sec:observer-on-flats}
adapts readily, if in part less explicitly, to counterparts of $M_0$
and $M_-$ in which $y$ is periodic and the horizon area finite. For
the $y$-periodized counterpart of~$M_0$,
the path-integral approach yields for the entropy one quarter of the
horizon area \cite{laf-flat-entropy}, while for the $y$-periodized
counterpart of $M_-$ the path-integral entropy contains the
additional factor~$\casehalf$.

\section{Summary and discussion}
\label{sec:discussion}

In this paper we have investigated thermal effects on the
$\RPthree$ geon and on a topologically analogous flat spacetime
$M_-$ via a Bogoliubov transformation, a particle detector, particle
emission and absorption coefficients, and stress-energy tensor
expectation values.  We fixed our attention to the 
Hartle-Hawking-like
vacuum $\hhvacgeon$ on the geon and to the Minkowski-like
vacuum $\minusvac$ on~$M_-$. We saw that, at finite times, these
states are not exactly thermal unless they are sampled with a probe
that couples to only half of the field modes. However, $\hhvacgeon$
becomes fully thermal, at the usual Hawking temperature, in the
distant past and future in the far
exterior Schwarzschild region, and
$\minusvac$ similarly becomes fully thermal at early and late
Rindler times in its Rindler wedge, with the usual Rindler
temperature for a uniformly accelerated observer. In addition, we
found some evidence for the thermality of $\hhvacgeon$ at the geon
spatial infinity, at arbitrary values of the exterior Schwarzschild
time.  In the case of the geon, some of these results rest on a set
of plausible assumptions about the asymptotic behavior of the
Hartle-Hawking Green's functions $G_{\rm K}(x,x')$ on the Kruskal
manifold when
$x$ and $x'$ are in opposite asymptotic regions, whereas for $M_-$
the asymptotic behavior of the relevant Green's function
could be directly verified. We have
also noted a discrepancy in the calculations for the geon entropy
via late-time thermodynamic arguments and the path-integral
method, and discussed some probable resolutions.

One may well ask whether it was necessary to perform each of these
calculations separately. As the Bogoliubov transformations contain
the full information about
$\hhvacgeon$ in terms of Boulware modes, and about $\minusvac$ in
terms of the Rindler modes, the detector response and the particle
emission-absorption probabilities must already be somehow encoded in
the Bogoliubov coefficients. Understanding this encoding would be
particularly useful for the geon, as one would then hope to use the
expressions (\ref{hhvacgeon-expanded}) and
(\ref{pac-hhvacgeon-expanded}) for the Boulware-mode content of
$\hhvacgeon$ to show that the detector response does indeed become
thermal in the asymptotic past or future. Unfortunately, only a
partial understanding of the encoding seems to be available
\cite{hu-matacz}.

What our results do strongly suggest 
is that an essential part of the
information in the Bogoliubov transformation resides in the phase
correlations between the alpha and beta
coefficients.\footnote{We thank Bei-Lok Hu for stressing this point
to us.}
We saw that the Boulware-mode occupation number expectation values do
not distinguish between the vacuum $\hhvacgeon$ on the geon and the
Hartle-Hawking vacuum $\hhvackrus$ on the Kruskal spacetime, despite
the fact that $\hhvackrus$ is static in the exterior region while
$\hhvacgeon$ is not. 
The number expectation values are determined
by the absolute values of the beta coefficients,
and these carry no information about the 
phases of the coefficients. 
To see explicitly where the phases enter, 
we observe that the geon $W$-modes
(\ref{W-modes-geon}) are not invariant, 
not even up to an overall phase, 
under exterior Schwarzschild time translations, 
because such translations would turn the
phases of the two terms in (\ref{W-modes-geon}) in the opposite
directions: 
these two terms in turn determine the alpha and beta
coefficients, and thus encode into the phases of the 
coefficients the fact that the spacetime has a distinguished value of
the exterior Schwarzschild time. In contrast, the $W$-modes on
Kruskal are invariant up to an overall phase under Schwarzschild time
translations, because the two terms in the Kruskal counterpart of
(\ref{W-modes-geon}) live in opposite exterior regions
\cite{unruh-magnum}. Analogous considerations hold for~$M_-$. One
might speculate on whether a continued study of the these spacetimes
would shed further light on the connection between Bogoliubov
coefficients and other measures of thermal behavior.

We have characterized $\hhvacgeon$ and $\minusvac$ as states that are
induced by well-studied states on the universal covering spacetimes,
as states with certain analytic properties, and as the no-particle
states for modes that are positive frequency with respect to the
horizon generator affine parameters. One might ask whether other,
perhaps better and more geometrical, characterizations of these
states could be given. One might also seek uniqueness theorems that
would single out these states, in the same way that $\hhvackrus$ is
selected on the Kruskal manifold or the Minkowski vacuum is selected
on Minkowski spacetime \cite{wald-qft}. For example, it might be
possible to adapt the Kay-Wald conditions \cite{kay-wald} to the geon
in a way in which it would suffice for the Killing vector to be only
local. On a converse note, one might seek to understand the late-time
thermal properties of $\hhvacgeon$ in the framework of general
late-time behavior in dynamical black hole spacetimes. It might for
example be possible to adapt the theorems of Fredenhagen and Haag
\cite{freden-haag} to initial conditions compatible with the geon.

Our analysis of the particle emission and absorption cross sections
relied in an essential way on the analytic structure of the Feynman
propagators in $\hhvacgeon$ and~$\minusvac$. Although the propagators
are `periodic in imaginary time' in a certain local sense, we saw
that this local periodicity is not associated with a globally-defined
$\Uone$ isometry of the Riemannian sections of the spacetimes. The
absence of such an isometry reflects the nonstaticity of $\hhvacgeon$
and~$\minusvac$, and would certainly cast doubts on simply
identifying the local period of the imaginary time as an inverse
temperature. A~similar argument in another context was made in Ref.\
\cite{bran-kahn}. It should therefore be emphasized that we used the
local periodicity in imaginary time only as a mathematical device in
the calculation of a genuinely Lorentzian observable quantity,
namely, the ratio of the emission and absorption cross sections at
late times. The thermal conclusion was drawn from this ratio.

For a class of operators that only couple to a judiciously-chosen
half of the field modes, the expectation values in $\hhvacgeon$ and
$\minusvac$ were seen to be thermal for all times, and not just in
the limit of early or late times. One may ask whether detectors with
such couplings could be built of matter whose underlying Lagrangian
has reasonable properties, including general covariance and, in the
case of~$M_-$, local Lorentz
invariance.\footnote{We thank Karel Kucha\v{r} for raising this
question.}
While it is likely that this can be achieved with sufficiently
complicated composite detectors, at least in an approximate sense
over a range of the field modes, we expect the answer to be negative
for detectors that couple locally to the field at finite positions and
times.

\begin{figure}
\begin{center}
\vglue 3 cm
\leavevmode
\epsfysize=8cm
\epsfbox{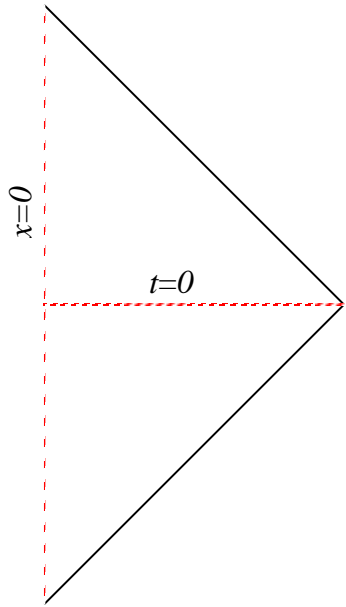}
\end{center}
\vskip 3 cm
\caption{%
A~conformal diagram of the constant $y$ and $z$ sections
of the spacetime~$M_-$.
When the diagram is understood to depict~$M_-$,
the region $x>0$ is identical to
that in the diagram of
Figure~\ref{fig:Mnought},
each point representing a suppressed cylinder.
At $x=0$,
each point in the diagram represents a suppressed open M\"obius
strip ($\simeq\RPtwo\setminus\{\hbox{point}\}$),
with the local coordinates $(y,z)$ identified by
$(y,z)\sim(-y,z+a)$.
Because of the suppressed
dimensions, the infinities of the diagram do not
faithfully represent the infinity structure of~$M_-$.}
\label{fig:Mminus}
\end{figure}

\begin{figure}
\begin{center}
\vglue 3 cm
\leavevmode
\epsfysize=6cm
\epsfbox{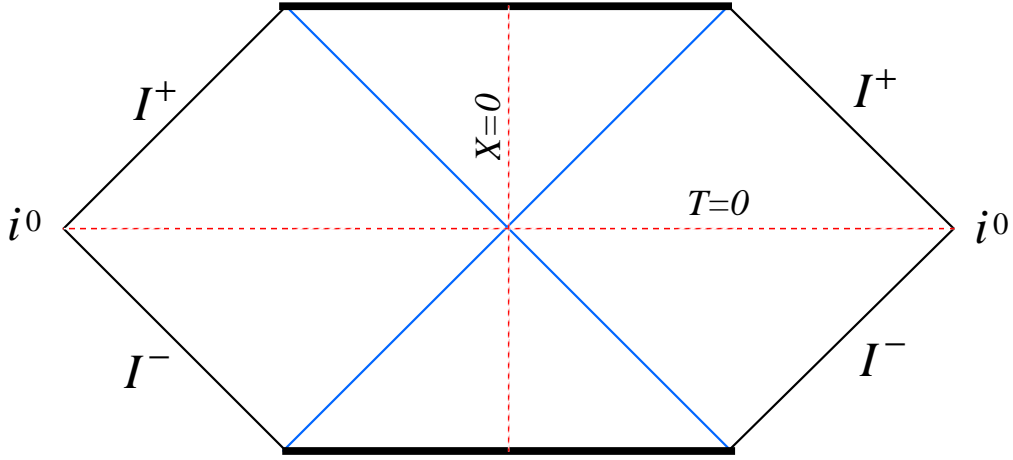}
\end{center}
\vskip 3 cm
\caption{%
A~conformal diagram of the Kruskal manifold.
Each point represents a suppressed $S^2$ orbit
of the $\Othree$ isometry group.
$(T,X)$ are the Kruskal coordinates introduced in
subsection~\ref{subsec:kruskal-def}, and 
the hypersurfaces $T=0$ and $X=0$ are shown.
The involution $J^L$ (\ref{JL-in-Kruskalcoords}) consists
of the reflection $(T,X) \mapsto (T,-X)$
about the vertical axis, followed by the
antipodal map
$(\theta,\varphi) \mapsto (\pi-\theta, \varphi + \pi)$
on the suppressed two-sphere.}
\label{fig:Kruskal}
\end{figure}

\begin{figure}
\begin{center}
\vglue 3 cm
\leavevmode
\epsfysize=6cm
\epsfbox{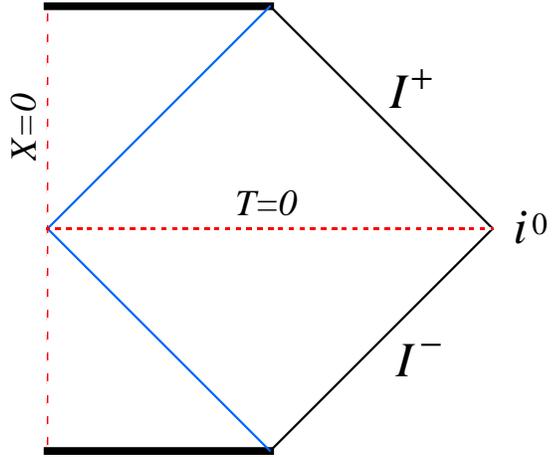}
\end{center}
\vskip 3 cm
\caption{%
A~conformal diagram of the
$\RPthree$ geon \protect\cite{topocen}.
Each point represents a suppressed orbit
of the $\Othree$ isometry group.
The region $X>0$ is isometric to the region
$X>0$ of the Kruskal spacetime,
shown in figure~\ref{fig:Kruskal};
in particular, the $\Othree$ isometry orbits
in this region are two-spheres.
At $X=0$, the $\Othree$ orbits have topology~$\RPtwo$.}
\label{fig:RPthree}
\end{figure}

\begin{figure}
\begin{center}
% \vglue 3 cm
\leavevmode
\epsfysize=14cm
\epsfbox{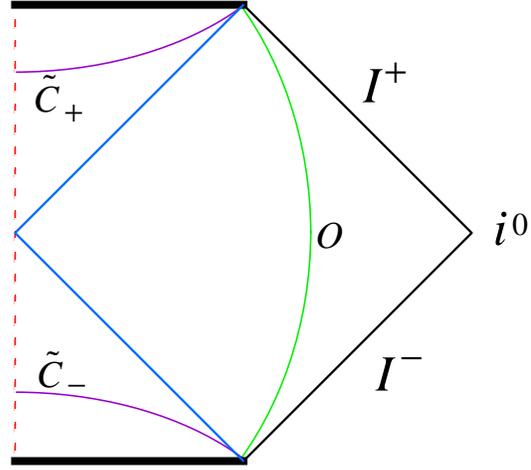}
\end{center}
% \vskip 3 cm
\caption{%
A conformal diagram for the emission and absorption calculation of
subsection \ref{subsec:rpthree-path} on the $\RPthree$ geon. The
timelike hypersurface $O$ and the spacelike hypersurfaces 
${\tilde C}_+$ and ${\tilde C}_-$ are shown.
The diagram for the corresponding
emission and absorption calculation on Kruskal
is shown in
Figure 3 of Ref.\ \protect\cite{hh-vacuum}.
}
\label{fig:RPthree-em-abs}
\end{figure}

For developing a geometrical understanding of the thermal properties
of our spacetimes, and for testing those conclusions that rested in
part on an unverified assumption about the Hartle-Hawking Green's
function on the Kruskal spacetime, it would be useful to have at hand
more examples of spacetimes with similar properties. One example of
interest is the spacetime $M_+:=M/J_+$, where $M$ is Minkowski
spacetime and the map $J_+$ reads, in the notation of
subsection~\ref{subsec:flat-spaces},
\begin{equation}
J_+: (t,x,y,z)  \mapsto  (t,-x, y,z+a)
\ \ .
\label{jplus}
\end{equation}
It is easily seen that the spacetime $M_0$ introduced in subsection
\ref{subsec:flat-spaces} is a double cover of~$M_+$, and that $M_+$
provides another
flat analogue of the $\RPthree$ geon, distinct from~$M_-$.
In fact, $M_+$ is in its isometry structure even more closely
analogous to the geon than~$M_-$: the two-dimensional conformal
diagram for $M_+$ is as in Figure~\ref{fig:Mminus}, but, unlike
for~$M_-$, the boundary of the diagram at $x=0$ now depicts a set in
$M_+$ that is geometrically distinguished in terms of the orbits of
the isometry group, in analogy with the boundary $X=0$ in the geon
diagram in Figure~\ref{fig:RPthree}. All our results for $M_-$ adapt
to~$M_+$, with 
conclusions that are qualitatively similar but 
exhibit some quantitative differences. 
In particular, as
$J_+$ leaves the coordinate $y$ invariant, the counterpart of
(\ref{minusvac-expanded-pac}) displays no correlations between
different values of~$y$, and the counterpart of (\ref{Delta-Gplus})
does not involve~$y_0$. $M_+$~is globally hyperbolic, but not space
orientable: its spatial topology is $\BbbR$ times the open M\"obius
strip. We have focused the present paper on $M_-$ in favor of $M_+$
in order to allay the suspicion that nonorientability might have been
a factor in the results.

Another spacetime with similar properties arises from taking the
quotient of de~Sitter space with respect to a $\BbbZ_2$ isometry
group in such a way that the spatial topology becomes~$\RPthree$.
Explicitly, if we realize de~Sitter space as the hyperboloid
\begin{equation}
\alpha^2 = - U^2 + V^2 + X^2 + Y^2 + Z^2
\end{equation}
in five-dimensional Minkowski space with the metric
\begin{equation}
ds^2 = - dU^2 + dV^2 + dX^2 + dY^2 + dZ^2
\ \ ,
\end{equation}
we can choose the nontrivial generator of the $\BbbZ_2$ 
to act as
\begin{equation}
(U, V, X, Y, Z) \mapsto (U, -V, -X, -Y, -Z)
\ \ .
\end{equation}
One would expect similar thermal results to hold for this spacetime.
It is in fact known that particle detectors in this spacetime behave
in the expected way \cite{schleich-priv}.

Our most intriguing result is probably the factor $\casehalf$ in the
path-integral-approach entropy of the geon, compared with the
Bekenstein-Hawking entropy of a Kruskal hole with the same mass.
While we argued that 
this result is mathematically understandable in view of the
complexified geometry of the geon, its physical significance, or
indeed physical correctness, remains
unclear. It should prove
interesting to see whether this factor $\casehalf$ might arise within
any state-counting approach to the geon entropy.

\acknowledgments
We would like to thank John Friedman for teaching us the geometry
of the $\RPthree$ geon and asking whether this spacetime has a
Hawking temperature.
We have also benefited from discussions and
correspondence with numerous other colleagues, including
Abhay Ashtekar,
Robert Brandenberger,
Andrew Chamblin,
Gary Gibbons,
Petr H\'aj\'{\i}\v{c}ek,
Gary Horowitz,
Bei-Lok Hu,
Ted Jacobson,
Karel Kucha\v{r},
Nicholas Phillips,
Alpan Raval,
Kristin Schleich,
Rafael Sorkin,
and
Bob Wald.
We thank the Banach Center of the Polish Academy of Science
for hospitality during the early stages of the work.
This work was supported in part by NSF grants
PHY94-21849
and
PHY97-22362,
and by research funds provided by Syracuse University. 
The work of one of us (J.L.) was done in part 
during a leave of absence from 
Department of Physics, 
University of Helsinki.

\newpage

\newpage

\newpage

\newpage

\end{document}